\documentclass[prb,aps,twocolumn,superscriptaddress]{revtex4-1}
\usepackage{graphicx}
\usepackage{amsmath,amssymb}
\begin{document}
\title{Uniform electron gas at finite temperatures}
\author{Travis Sjostrom}
\affiliation{Theoretical Division, Los Alamos National Laboratory,
Los Alamos, New Mexico 87545}
\author{James Dufty}
\affiliation{Department of Physics, University of Florida,
Gainesville Florida 32611}
\date{August 8, 2013}

\begin{abstract}
We calculate the free energy of the quantum uniform electron gas for
temperatures from near zero to 100 times the Fermi energy,
approaching the classical limit. An extension of the Vashista-Singwi 
theory to finite temperatures and self-consistent compressibility 
sum rule is presented. Comparisons
are made to other local field correction methods, as well as recent
quantum Monte Carlo simulation and classical map based results.
Accurate fits to the exchange-correlation free energy from both
theory and simulation are given for future practical applications.
\end{abstract}
\maketitle

\section{Introduction}

The uniform, or homogeneous, electron gas (UEG), also known as jellium or as
a one component plasma, is a well-studied system in physics. It is important
as a proving ground for method development. Accurate results provide a
better understanding of the rich underlying physics of classical and quantum
Coulomb correlations, as well as provide a basis for approximations in more
complicated real systems. One important case of note is density functional
theory at zero temperature, in which local density approximations (LDA)
using the UEG results for the exchange and correlation (XC) energy have
proved remarkably successful in systems as diverse as molecules to exotic
phases of highly compressed matter. A challenge in current research is
simulations of warm dense matter (WDM), motivating pursuit of accurate
finite temperature UEG results for the corresponding development of
temperature dependent functionals.

The zero temperature UEG was the subject of much theoretical development in
60's and 70's of the last century. RPA and beyond RPA dielectric
approximations were particularly successful \cite%
{Lindhard,Gell-MannBrueckner,STLS,VS} in appropriate limits. However in 1980
Ceperley and Alder \cite{CA80} produced benchmark quantum Monte Carlo (QMC)
results with nearly exact accuracy across a wide range of densities, though
the fixed-node approximation does lead to small errors in the results for
high densities. These accurate values for the UEG XC energy provided the
essential LDA needed for designing functionals beyond LDA \cite{PZ82}.
Almost all subsequent zero temperature DFT formulations make use of this LDA
obtained from the UEG simulation in some explicit way. The corresponding LDA
for development of finite temperature DFT, firmly based in the finite
temperature UEG, has been lacking up until now.

There has been much less development for the finite temperature UEG,
in part due to lack of experimental motivation. Now, experimental
conditions of WDM span the range from zero temperature to far above
the Fermi temperature. Until very recently \cite{Brownetal} there
has not been any QMC type simulations in this range to extend those
of Ceperley and Alder at zero temperature. RPA calculations were
done originally by Gupta and Rajagopal \cite{GuptaRajagopal}, and
later revised and fits provided by Perrot and Dharma-wardana
\cite{PDw84}. Shortly after, beyond RPA calculations were
done including static \cite{TanakaIchimaru,DAC} and dynamic \cite%
{SchwengBohm} local field corrections. A finite temperature Vashista-Singwi
type calculation (VS) was done using an approximate form for the local field
corrections \cite{StolzmannRosler}. In addition other methods have been
proposed including the so-called modified convolution approximation \cite%
{TanakaIchimaruMCA} and interpolation approximations \cite{Ebeling}.
Most recently, methods of mapping the quantum problem to a
corresponding classical system have been proposed
\cite{PDw2000,DuttaDufty}, where effective classical strong coupling
methods such as molecular dynamics simulation and liquid state
theory can be applied \cite{Hansenbook}. Further details of some of
these theories are given in the results and comparisons sections.

Two thermodynamic parameters are required to describe the equilibrium UEG,
chosen here to be the density $n$ and temperature $T$ . When measured
relative to the Fermi temperature, the dimensionless temperature is%
\begin{equation}
t\equiv k_{B}T/E_{F},\;
\end{equation}%
where $E_{F}=\hbar ^{2}q_{F}^{2}/2m_{e}$ is the Fermi energy, and $%
q_{F}=(3\pi ^{2}n)^{1/3}$ is the Fermi wave vector. The density is typically
specified in terms of the electron Wigner-Seitz length $r_{0}=(4\pi
n/3)^{-1/3}.$ When measured relative to the Bohr radius $a_{B}=\hbar
^{2}/(m_{e}e^{2})$ its dimensionless form is%
\begin{equation}
r_{s}\equiv r_{0}/a_{B}.
\end{equation}%
Dimensionless thermodynamic properties therefore can be expressed as
functions of $t,r_{s}$. The importance of Coulomb coupling is measured by a
coupling constant defined as the ratio of the Coulomb energy for a pair at
the distance $r_{0}$ relative to the kinetic energy per particle. In the
classical limit the appropriate kinetic energy is $k_{B}T$ and the classical
coupling constant is
\begin{equation}
\Gamma \equiv e^{2}/(r_{0}k_{B}T)\;.
\end{equation}%
It is related to $r_{s},t$ by $\Gamma =2\lambda ^{2}r_{s}/t$, where $\lambda
=(4/9\pi )^{1/3}$. At very low temperatures the relevant kinetic energy is $%
E_{F}$ and the corresponding coupling constant at $t=0$ is a function of $%
r_{s}$ only.

As previously noted the $t=0$ limit has seen much development culminating in
high accuracy ab initio simulations. This has also been the case for the
classical limit $t\gg 1$ \cite{Hansen}. The theoretical development in the
intermediate Fermi-degeneracy region mentioned above has not been
benchmarked so that the relative accuracy of the various methods is unknown.
The objective here is to present a improvement of the finite temperature
Vashista-Singwi model by including a consistency requirement on the
dielectric function and the pressure derived from it (the exact
compressibility sum rule for the small wave vector limit of the dielectric
function). The temperature dependence of the structure (pair correlation
function) and thermodynamics (free energy, compressibility) are calculated
from this improved Vashista-Singwi approximation (VSa) in the range $0\leq
t\leq 10$ for a wide range of $r_{s}$ corresponding to WDM conditions. 
The approach of the free energy to the classical limit is explored 
also at much higher temperatures.
Comparisons with several other theoretical models and the new QMC simulation
results are also given. In this way, some assessment of the quality and
trends of the results is established.

\section{Vashista-Singwi model with self-consistent compressibility}

We calculate the UEG at finite temperature by means of an approximate
dielectric function of the form
\begin{equation}
\varepsilon (\mathbf{q},\omega )=1-\frac{v_{q}\chi _{0}(\mathbf{q},\omega )}{%
1+G(\mathbf{q})v_{q}\chi _{0}(\mathbf{q},\omega )}  \label{eq:dielc}
\end{equation}%
where $v_{q}=4\pi e^{2}/q^{2}$ is the Coulomb potential and $\chi _{0}(%
\mathbf{q},\omega )$ is the finite temperature polarizability of the
non-interacting UEG, and $G(\mathbf{q})$ is the static local field
correction (LFC). For simplicity of notation, the dependence of these
functions on $r_{s},t$ is not made explicit except where needed for clarity
or emphasis.

The static structure factor is found by the fluctuation-dissipation
theorem as a sum over the Matsubara frequencies for the
polarizabilities of the interacting system \cite{TanakaIchimaru} as
\begin{align}
S(\mathbf{q})& =-(\beta n)^{-1}\sum_{l=\infty }^{\infty }\frac{1}{v_{q}}%
\left( \frac{1}{\varepsilon (\mathbf{q},z_{l})}-1\right)  \notag \\
& =-(\beta n)^{-1}\sum_{l=\infty }^{\infty }\frac{\chi _{0}(\mathbf{q},z_{l})%
}{1-[1-G(\mathbf{q})]v_{q}\chi _{0}(\mathbf{q},z_{l})}  \label{eq:Sq}
\end{align}%
where $z_{l}=2\pi il/\beta \hbar $ and in the second line we have made the
static LFC approximation consistent with Eq. \ref{eq:dielc}.

We choose for $G(\mathbf{q})$ the form given originally by Vashista and
Singwi (VS) \cite{VS} in the following temperature dependent generalization
\begin{align}
G(\mathbf{q})=& \left( 1+a(r_{s},t)n\frac{\partial }{\partial n}\right)
\times  \notag \\
& \left( -\frac{1}{n}\int {\ \frac{d\mathbf{q}^{\prime }}{(2\pi )^{3}}\frac{%
\mathbf{q\cdot q^{\prime }}}{q^{\prime 2}}\left[ S(\mathbf{q}-\mathbf{%
q^{\prime },}r_{s},t)-1\right] }\right)  \label{eq:Gq}
\end{align}%
where $a(r_{s},t)$ is a parameter determined below. Contained within this
form are the LFC for other finite temperature calculations. For example, $%
G=0 $ is RPA and $a=0$ is the finite temperature STLS approximation. In the
original introduction by VS $a$ was taken as a constant equal to $2/3$ for
the zero temperature UEG. This value was chosen to provide better agreement
with the compressibility sum rule (below) for metallic densities, with
discrepancies only becoming noticeable around $r_{s}=4$.

For a given value of $a$, Eqs. \ref{eq:Sq} and \ref{eq:Gq} form a coupled
pair of equations that must be solved self-consistently. The resulting $S$
and $G$ may then be used to calculate the dielectric function and other
properties of the UEG. The compressibility sum rule (CSR) is an exact
property of the UEG given by
\begin{equation}
\lim_{\mathbf{q}\rightarrow 0}\varepsilon (\mathbf{q},\omega
=0)=1+v_{q}n^{2}\kappa
\end{equation}%
where $\kappa $ is the thermodynamic compressibility defined in terms of the
pressure by
\begin{equation}
\frac{1}{\kappa }=n\frac{\partial P}{\partial n}\;.  \label{eq:kappaEOS}
\end{equation}%
Calculation of the compressibility from an approximate dielectric function
will generally result in a different value than that obtained from the
derivative of the associated pressure. In order to enforce consistency of
the pressure and dielectric forms we define $a(r_{s},t)$ for satisfaction at
all $r_{s}$ and $t$. The two expressions for the compressibility can be
written in the equivalent form%
\begin{equation}
\frac{\kappa _{0}}{\kappa }=1+\kappa _{0}n^{2}\frac{\partial
^{2}(nf_{xc}(n,t))}{\partial n^{2}}=1-\kappa _{0}n^{2}\gamma 4\pi e^{2},
\label{eq:kapparatio}
\end{equation}%
where $\kappa _{0}$ is the compressibility for the non-interacting UEG. In
the first equality the pressure has been expressed in terms of
exchange-correlation free energy per particle $f_{xc}$. In the second
equality the dielectric function in the form of Eq. \ref{eq:dielc} has been
used, with the definition $\gamma \equiv \lim_{q\rightarrow 0}q^{-2}G(q)$.

In order to calculate $f_{xc}$ we perform an integration of the
interaction energy over the Coulomb coupling constant. In the
following this integration is replaced as an integration over
$r_{s}$ at constant $t$ \cite{SchwengBohm}
\begin{equation}
f_{xc}(r_{s},t)=\frac{1}{r_{s}^{2}}\int_{0}^{r_{s}}dr_{s}^{\prime
}\;r_{s}^{\prime }e_{\mathrm{int}}(r_{s}^{\prime },t).
\label{eq:fxc1}
\end{equation}
Here $e_{\mathrm{int}}(r_{s},t)$ is the average interaction energy
per particle (average Coulomb potential energy), as distinct from
the corresponding exchange-correlation energy. This average
interaction energy can be expressed in terms of the structure factor
$S(\mathbf{q})$ for evaluation from the
above dielectric theories.  From this point on the reduced wave vector $%
x=q/q_{F}$ and Hartree atomic units ($\hbar =m_{e}=e=1$) are used.
The XC free energy per particle is then given by
\begin{equation}
f_{xc}(r_s,t)=\frac{1}{\pi \lambda
r_{s}^{2}}\int_{0}^{r_{s}}dr_{s}^{\prime }\int_{0}^{\infty }dx\left[
S(x)-1\right] .  \label{eq:fxc}
\end{equation}

For numerical evaluation Eqs. \ref{eq:Sq} and \ref{eq:Gq} are written in the
following forms.
\begin{equation}
S(x)=\frac{3}{2}t\sum_{l=-\infty }^{\infty }\frac{\Phi (x,l)}{1+(2\Gamma
t/\pi \lambda x^{2})[1-G(x)]\Phi (x,l)}
\end{equation}%
where
\begin{align}
\Phi (x,l)& =-\frac{\pi ^{2}}{q_{F}}\chi _{0}(q,z_{l})  \notag \\
& =\frac{1}{2x}\int_{0}^{\infty }\frac{dy\;y}{e^{y^{2}/t-\eta }+1}\ln {%
\left\vert \frac{(2\pi lt)^{2}+(x^{2}+2xy)^{2}}{(2\pi lt)^{2}+(x^{2}-2xy)^{2}%
}\right\vert }
\end{align}%
is the dimensionless free-electron polarizability. Here $\eta =\beta \mu
_{0} $ where $\mu _{0}$ is the chemical potential of the non-interacting
system, which may be found from $t$ through the Fermi integral
\begin{equation}
I_{1/2}(\eta )=\frac{2}{3}t^{-3/2}\;.
\end{equation}%
Additionally we use the form for the short wave length regime given in Ref. %
\onlinecite{TanakaIchimaru} (their Eq. 27) for the evaluation of $S(x)$.

Equation \ref{eq:Gq} is then given by
\begin{equation}
G(x)=G_{I}(x)+a(r_{s},t)\left( -\frac{x}{3}\frac{\partial }{\partial x}-%
\frac{r_{s}}{3}\frac{\partial }{\partial r_{s}}\right) G_{I}(x)
\end{equation}%
where
\begin{equation}
G_{I}(x)=-\frac{3}{4}\int_{0}^{\infty }y^{2}\left[ S(y)-1\right] \left( 1+%
\frac{x^{2}-y^{2}}{2xy}\ln {\left\vert \frac{x+y}{x-y}\right\vert }\right)
dy\;.  \notag
\end{equation}%
In practice, the derivatives with respect to $x$ and $r_{s}$ are taken by
finite difference approximations. For $x$ this is simply done for a
calculation at any $r_{s}$ and $t$. However, for $r_{s}$ this requires
having $G(x)$ for neighboring $r_{s}$ so we solve self-consistently five
points simultaneously $[r_{s}-2\delta ,r_{s}-\delta ,r_{s},r_{s}+\delta
,rs+2\delta ]$. The derivative and second derivative of the central point is
solved using finite differences and the derivatives of the neighboring
points are given by Taylor expansion about the central point using its
second derivative.

Beyond the self-consistency for $S$ and $G$ we impose self-consistency of
Eqs. \ref{eq:kapparatio} and \ref{eq:fxc} to find $a(r_{s},t)$. Figure \ref%
{fig:csr} shows the results for $t=1$ as a function of $r_s$. The top panel shows the
compressibility ratio calculated from the EOS (first equality of Eq. \ref%
{eq:kapparatio}) in comparison with that calculated from the dielectric
function (second equality of Eq. \ref{eq:kapparatio}). Several choices for $G$
are illustrated: RPA ($G=0$), STLS ($a=0$), the $t=0$ VS0 ($a=2/3$), and the
CSR constrained result here VSa ($a(r_{s},t)$). All of the methods produce
nearly identical results for the EOS calculations and are shown by the
single curve labeled EOS. Clearly the only results that give satisfaction of
the CSR is our curve where it is enforced. The lower panel shows the
self-consistent value of $a(r_{s},t=1)$ as a function of $r_{s}$.
Qualitatively similar results are obtained at other temperatures as well.
\begin{figure}[tbp]
\includegraphics[angle=-90,width=2.8in]{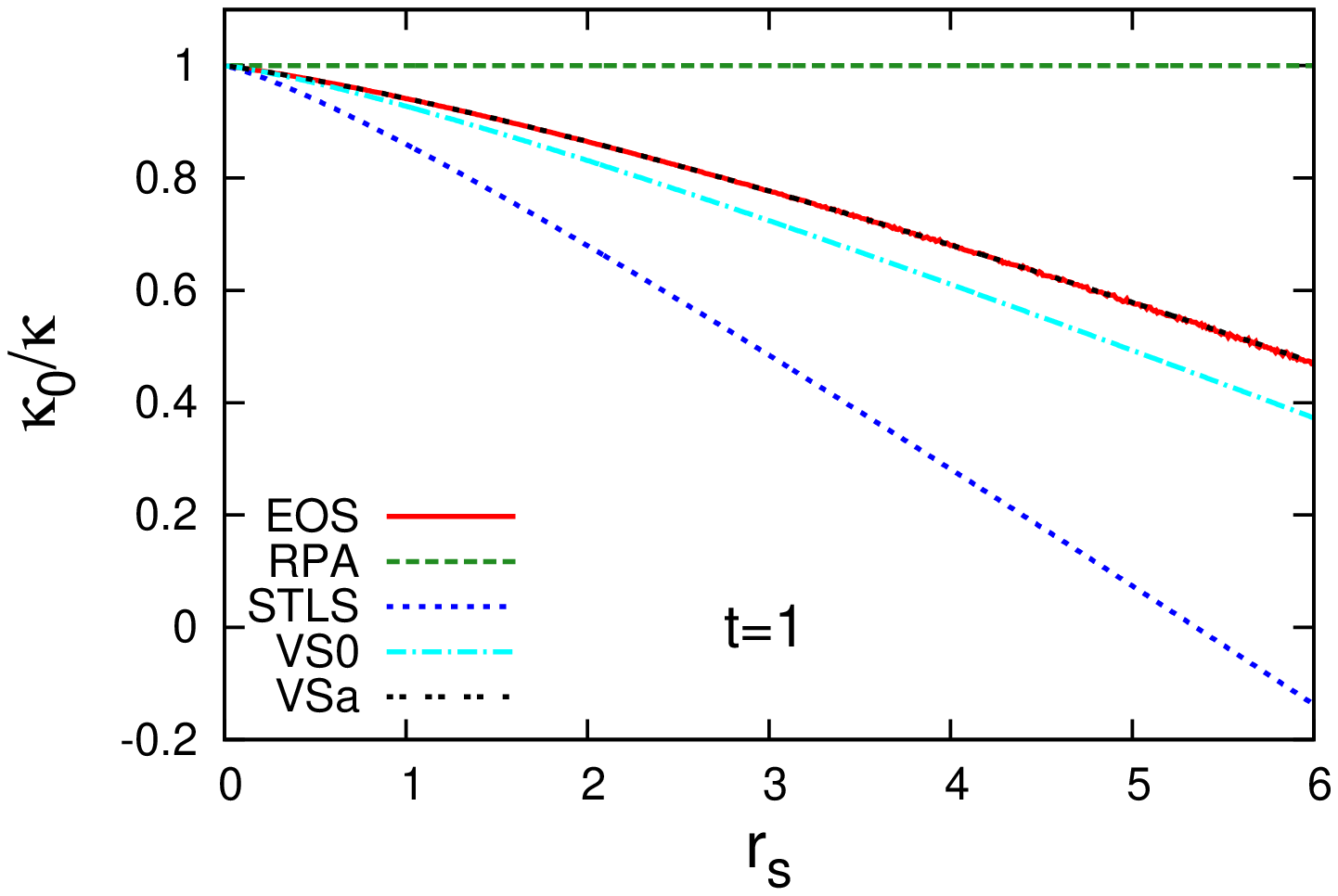} %
\includegraphics[angle=-90,width=2.8in]{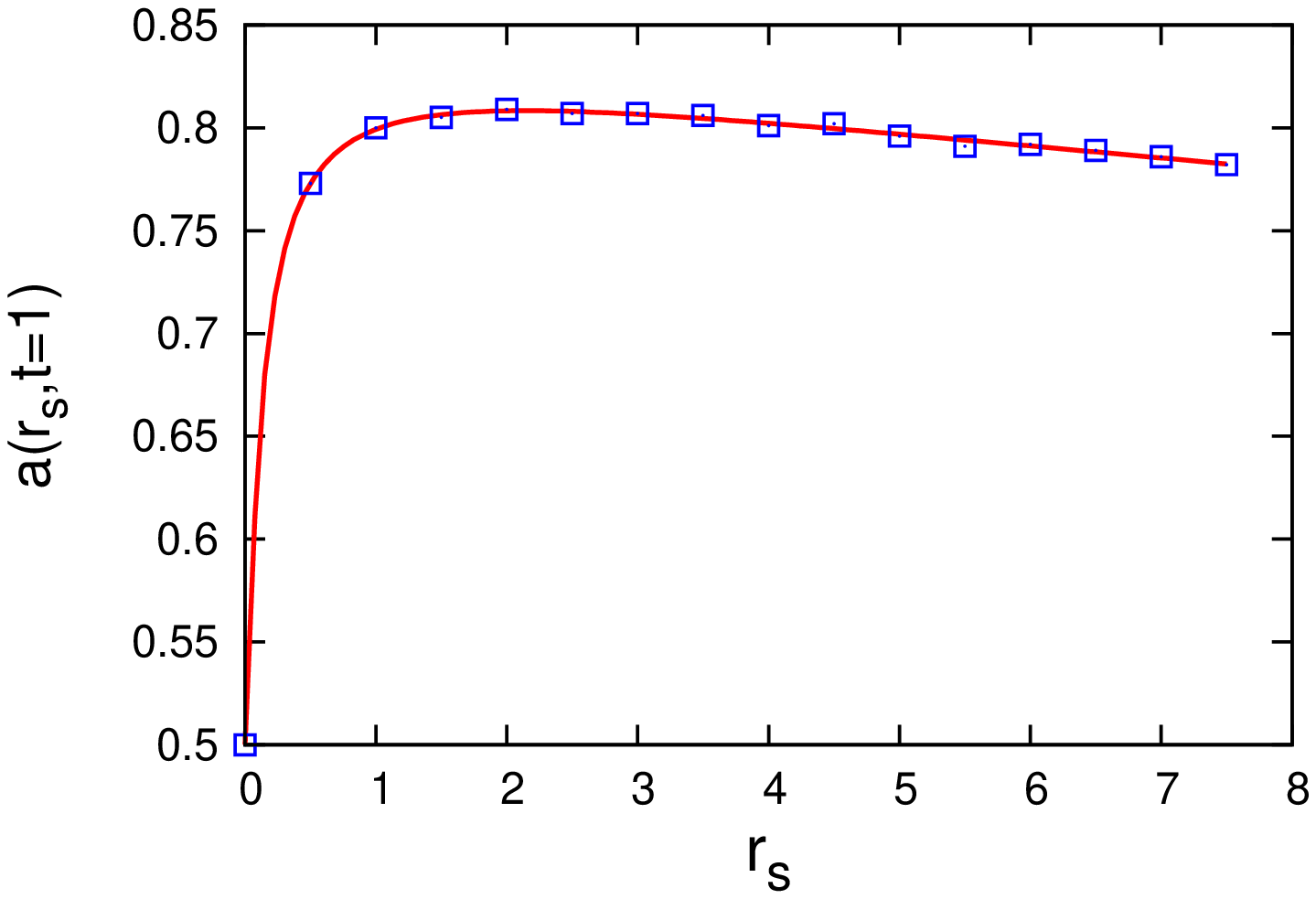}
\caption{Upper: Compressibility ratio from sum rule is plotted for various
approximate dielectric functions, along with ratio from equation of state.
Lower: The self consistent $a(r_{s},t)$ which satisfies CSR at $t=1$.}
\label{fig:csr}
\end{figure}

We perform the self-consistent calculation for $a(r_{s},t)$ over the
temperature and density plane at the values $%
t=[0.0625,0.125,0.25,0.5,1.0,1.5,2,3,4,6,8,10]$ and $r_{s}$ at integer and
half-integer values from $0-10$. For $r_{s}$, however, the self-consistent
calculation requires a fit of $a(r_{s},t)$ for all $r_{s}$ and so those
calculations for a given $t$ are performed for $r_{s}$ at $0.01$ spacing
from $0-10$. Integration for $S(x)$ and $G(x)$ are done up to $x=240$, the
Matsubara frequencies are summed up to $|l|=1000$.

\section{Results and comparisons}

Before presenting the results of our calculations we provide a brief list of
other methods and note those used for comparisons here. First we consider
the dielectric models described above, RPA, STLS, and VSa (the present
work). The above method for calculation is applied to all three, and it is
confirmed that the RPA and STLS results are in agreement with those provided
in the original studies \cite{GuptaRajagopal,PDw84,TanakaIchimaru}. STLS was
extended to include dynamic LFC in the ``quantum'' QSTLS method \cite%
{SchwengBohm}, though the QSTLS shows negligible energy differences 
with STLS for $t>1$. The modified convolution approximation, 
MCA, makes use of a static LFC, but solves a different set of integral
equations for $S$ and $G$ \cite{TanakaIchimaruMCA}. Interpolative
Pad\'e fits for high density, low density, and classical limits are
given by Ebeling \cite{Ebeling} and Kremp et al \cite{Krempetal}. A
quite different approach attempts to apply classical strong coupling
methods to the UEG using a quantum modified Coulomb potential and
effective thermodynamic parameters. The classical-map
hypernetted-chain method, CHNC, maps a quantum system with
temperature $T$, to a classical system with temperature $T_{cf}$ for
which classical calculations of correlation energy, pair
distribution functions, etc. are taken for the quantum system
\cite{PDw2000}. Another classical map, CM, enforces the equivalence
of the grand potential and two of its derivatives between a quantum
and classical system \cite{DuttaDufty}. Finally, restricted path
integral Monte Carlo simulation results have been performed over the
temperature and density range of interest \cite{Brownetal}. The
presentation below will compare our VSa results with other
dielectric models (RPA, STLS), classical map (CHNC), and quantum
simulations (RPIMC). Also shown for reference is the classical Monte
Carlo (CMC) simulations \cite{Hansen}.

\subsection{Interaction Energy}

Two equivalent expressions for the XC free energy are given in Eqs. \ref%
{eq:fxc1} and \ref{eq:fxc} as integrations over the coupling
constant (converted to $r_{s}$) of the interaction energy or
structure factor, respectively. The latter is convenient for
evaluation of the theories above, but the former is useful for
analysis of the results provided by RPIMC. In RPIMC the primary
results are the total average kinetic $k$ and average
potential energies $v$, which give the total internal energy $e_{\mathrm{tot}%
}=k+v$ (small casing indicates per particle). $v$ is in fact the
interaction energy $e_{\mathrm{int}}$ that appears in Eq. \ref{eq:fxc1}.
This is different from the XC energy, $e_{xc}$, whose RPIMC values
are the basis for the
numerical fit which is provided in Ref. \onlinecite{Brownetal}: $e_{xc}=e_{%
\mathrm{tot}}-e_{0}$, where $e_{0}=k_{0}$ is the ideal gas kinetic
energy. The XC energy is related to the XC free energy by the
thermodynamic identity $e_{xc}=f_{xc}+Ts_{xc}$ where $s_{xc}$ is the
excess entropy. Here we prefer to work with
$e_{\mathrm{int}}(r_{s},t)$, also provided in the RPIMC results
\cite{Brownetal}.

To facilitate the comparison of theory and simulation, we have first
fit the
RPIMC interaction energy data (see Appendix). The corresponding $e_{%
\mathrm{int}}(r_{s},t)$ from theory is obtained by a comparison of Eqs. \ref%
{eq:fxc1} and \ref{eq:fxc} for the identification
\begin{equation}
e_{\mathrm{int}}(r_{s},t)=\frac{1}{\pi \lambda r_{s}}\int_{0}^{\infty }dx\;%
\left[ S(x)-1\right] |_{t,r_{s}}.
\end{equation}%
The numerical fit using STLS has been given in Ref.
\onlinecite{Ichimaru93}; the corresponding fit using VSa is given here
(Appendix). Next, these fits are used in  Eq. \ref{eq:fxc1} to obtain
the XC free energy $f_{xc}$ for RPIMC, STLS, and VSa. Existing fits
for $f_{xc}$ from CHNC and CMC are also considered in the following.

We stress the importance of fits for $f_{xc}$ value for finite
temperature DFT and other applications, rather than those for
$e_{xc}$. It is the former that is required for the finite
temperature local density approximation in the construction of XC
functionals. The remainder of this paper continues analysis of the
various methods for cross validation and assessment of the best
approximation to be used.

\begin{figure}
\includegraphics[angle=-90,width=3.0in]{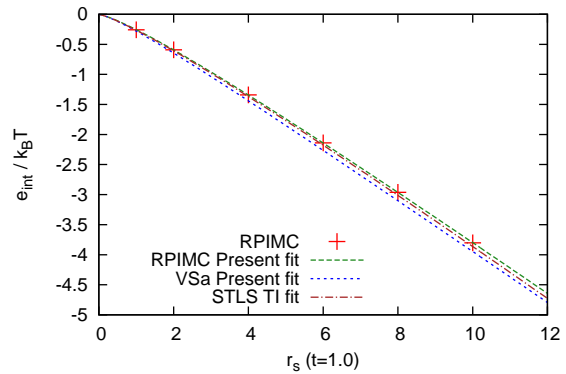} %
\caption{Interaction energy from RPIMC data and fits for RPIMC, VSa, and STLS as given in Appendix.}
\label{fig:eint}
\end{figure}

The interaction energy per particle divided by the temperature is
directly compared in Fig. \ref{fig:eint} for the RPIMC, VSa, and
STLS at $t=1$. The trends seen here hold for all $t$: the fit for
RPIMC is a very accurate representation of the raw data; the finite
temperature STLS is a very good approximation to the RPIMC, while
its ``improved'' version VSa is also good but with a larger
discrepancy from RPIMC.

\subsection{Equation of State}

The free energy is defined here as the sum of the non-interacting free
energy and the exchange-correlation free energy $F=F_{0}+F_{xc}$, where we
will consider the free energy per particle $F/N=f$. The non-interacting free
energy per particle is given by
\begin{equation}
f_{0}=F_{0}/N=-\frac{2}{3\beta }\frac{I_{3/2}(\eta )}{I_{1/2}(\eta )}+\frac{%
\eta }{\beta }
\end{equation}%
The exchange-correlation free energy per particle, $f_{xc}$, for this work
is given by Eq. \ref{eq:fxc}. Similarly the pressure is $P=P_{0}+P_{xc}$ and
found from the derivative of the free energy per particle for the
components, $P=n^{2}\;d(n(f_{0}+f_{xc}))/dn$. Additionally one may separate $%
f_{xc}$ into exchange only (X) and correlation (C) components using the
known value for $f_{x}$
\begin{equation}
f_{x}=-\frac{1}{2\pi }\left( \frac{\beta }{2}\right) ^{1/2}\frac{%
\int_{-\infty }^{\eta }[I_{1/2}(x)]^{2}\;dx}{I_{1/2}(\eta )}\;,
\label{eq:fx}
\end{equation}%
leaving the correlation component as the only value to calculate.
However, direct evaluation of Eq. \ref{eq:fxc} provides the XC
contribution
as a single term and fits are usually given for XC, so we plot in Fig. \ref%
{fig:fxc1} the XC free energy per particle, $f_{xc}$, relative to
the XC energy at zero temperature (known from zero temperature
quantum Monte Carlo calculations). The classical, high temperature
Debye-H\"{u}ckel limit has no exchange contribution and the
correlation component to first order is $f_{c}=-\frac{1}{3}\lambda
_{D}+\cdots$,
with $\lambda _{D}=(4\pi n\beta )^{1/2}.$ However encompassing this
limit are the CMC results of Hansen \cite{Hansen} which are shown
(Ref. \onlinecite{Hansen} also  provides quantum corrections but only
the classical excess free energy is shown here).

\begin{figure}[tbp]
\includegraphics[angle=-90,width=3.0in]{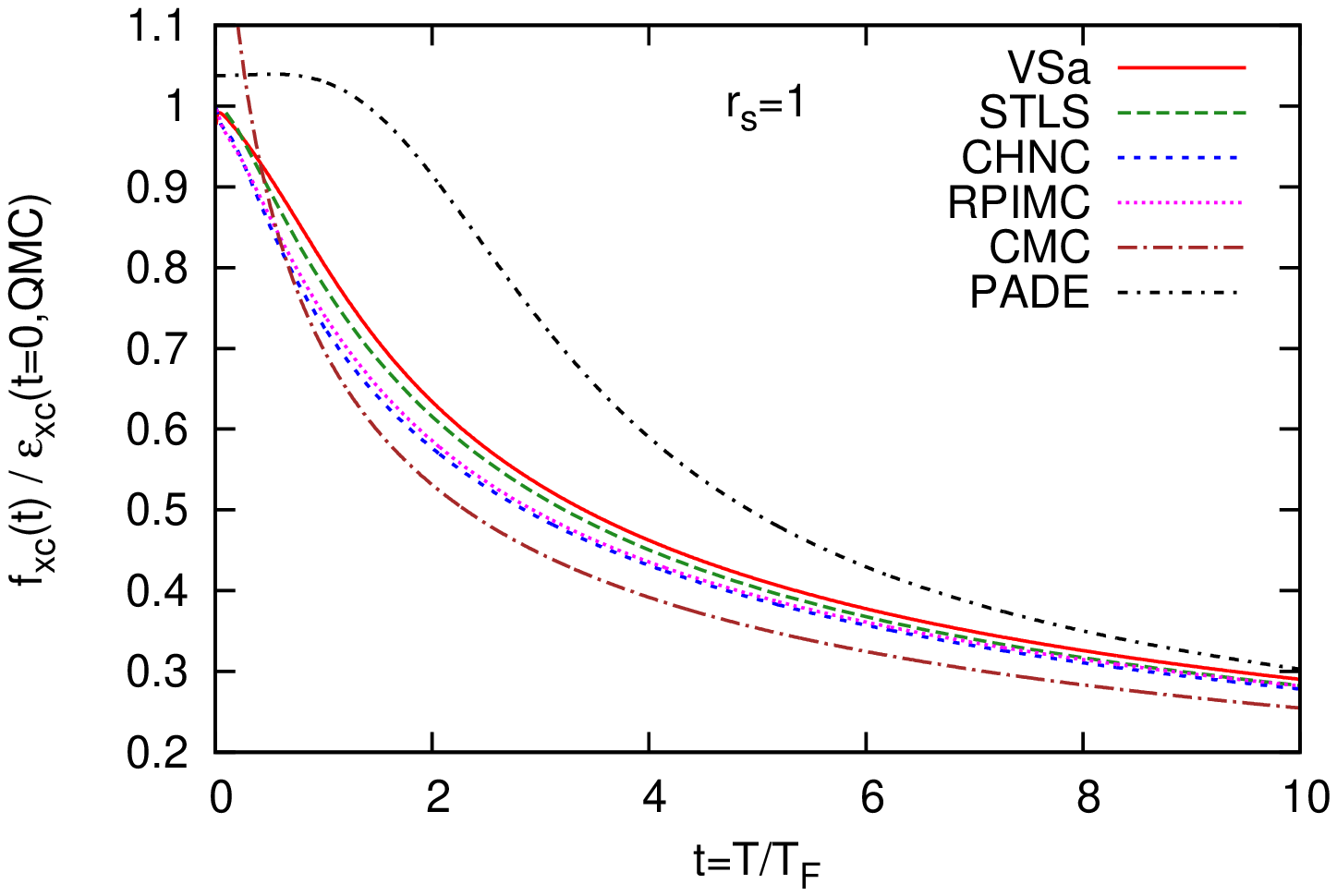} %
\includegraphics[angle=-90,width=3.0in]{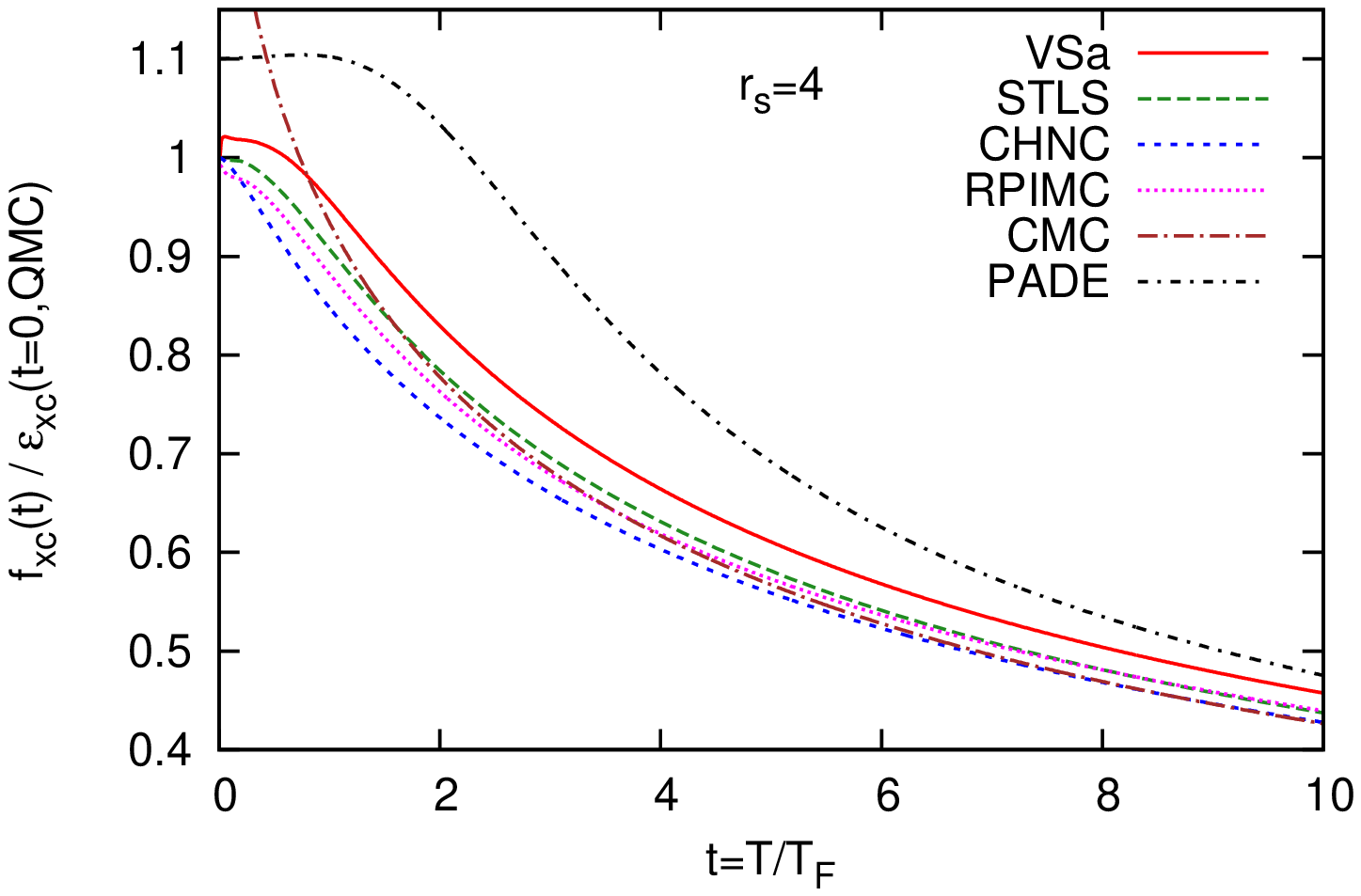}
\caption{XC free energy for several calculations, with the classical limit
plotted for comparison, relative to the known zero T XC energy.}
\label{fig:fxc1}
\end{figure}

In Fig. \ref{fig:fxc1}, we note first that there is a significant
temperature dependence predicted by all models for both $r_{s}=1$
and $4$ over the whole range considered $0\leq t\leq 10$.  Our VSa
results lie between those of RPA (not shown) and STLS. This trend
holds true for other properties such as $G(q)$, $S(q)$, and $g(r)$
as well. The CHNC (using the fit provided in Ref. \onlinecite{PDw2000}) is
systematically below these; like STLS it is a better approximation
to RPIMC than VSa, although all are quite similar. All of the
methods appear to be approaching the classical limit in the same
manor. The outlier is the Pad\'e interpolation due in major part to
the low $t$ limit being constructed to go to the Gell-Mann Brueckner
limit as opposed to the exact limit for larger $r_s$.

\subsection{Pair Correlation Function}

The pair correlation function $g(r)$ is calculated from the static structure
factor by
\begin{equation}
g(r)=1+\frac{3}{2r}\int_{0}^{\infty }x\sin (xr)[S(x)-1]dx
\end{equation}%
where $r$ is in units of $q_{F}^{-1}$.

\begin{figure}[t]
\includegraphics[angle=-90,width=1.65in]{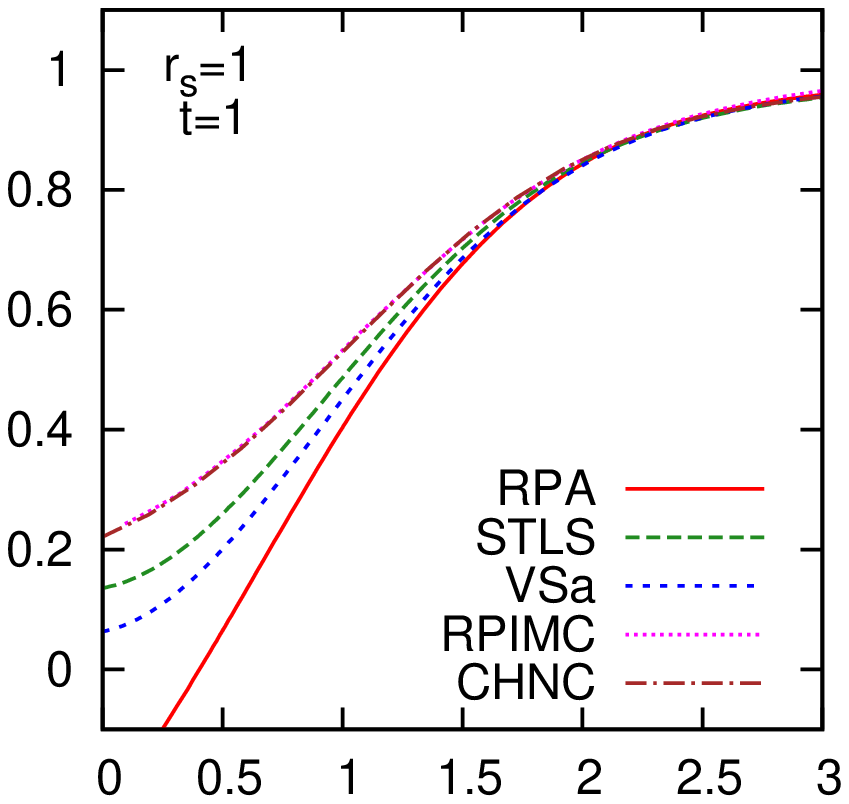} %
\includegraphics[angle=-90,width=1.65in]{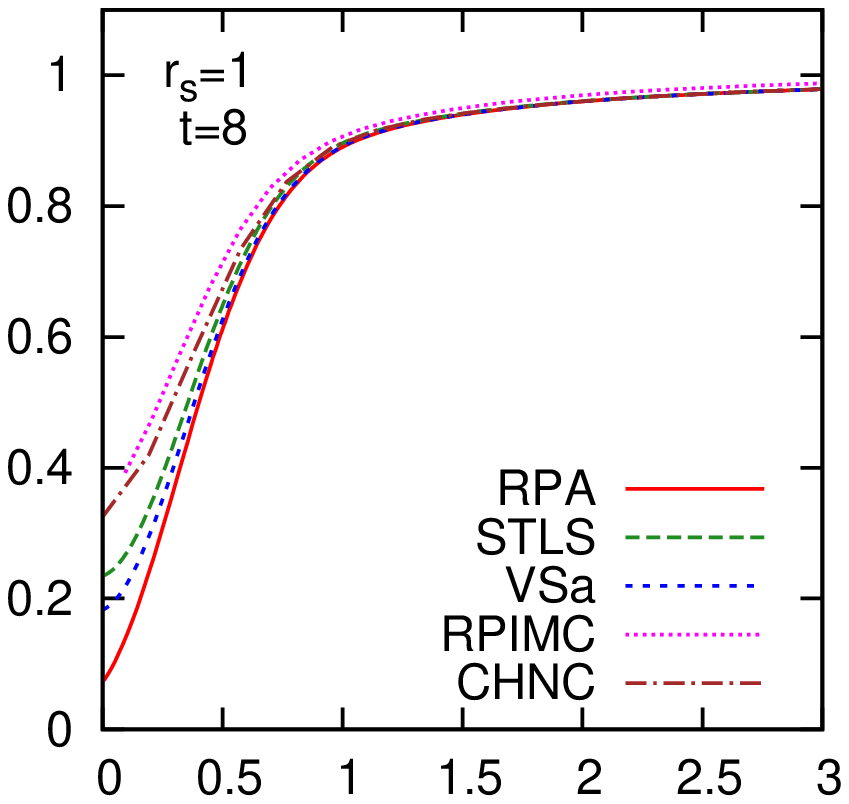} %
\includegraphics[angle=-90,width=1.65in]{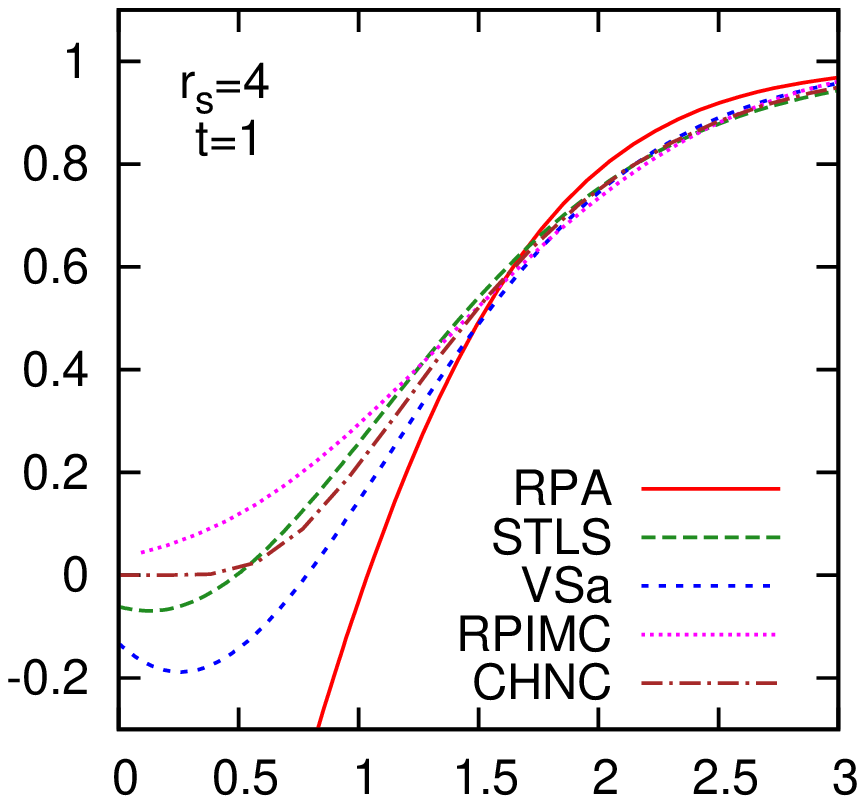} %
\includegraphics[angle=-90,width=1.65in]{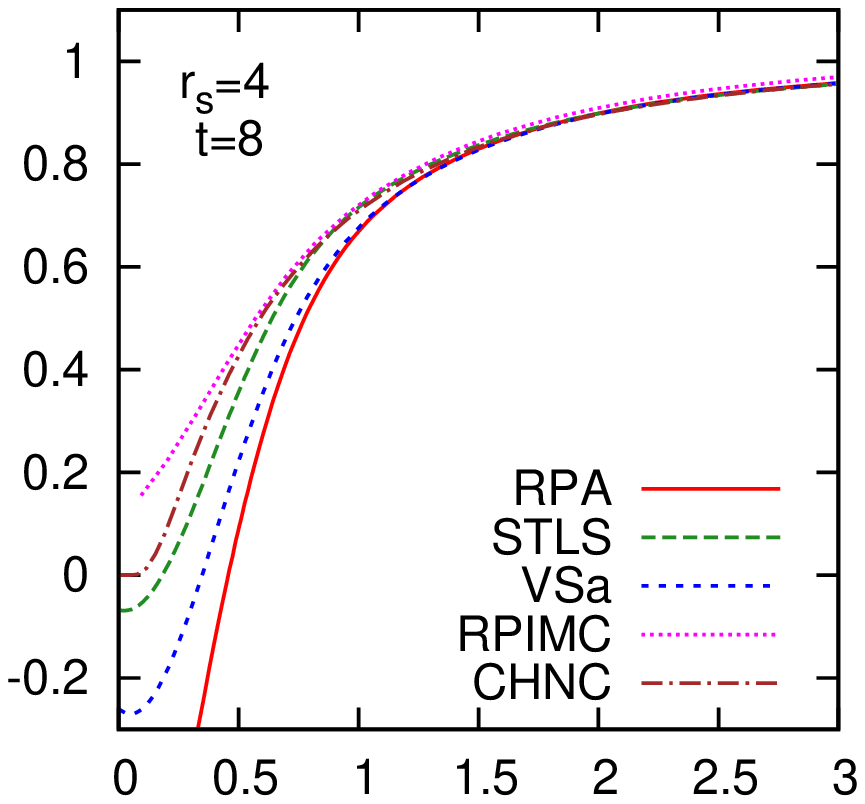}
\caption{Pair correlation functions at given $t$ and $r_s$. The $y$ and $x$
axis are $g(r)$ and $r$ respectively with $r$ in units of $q_F^{-1}$.}
\label{fig:gr1}
\end{figure}

The approximate dielectric methods are compared with RPIMC and the
classical map of Perrot and Dharma-wardana CHNC, in Fig.
\ref{fig:gr1}. Another classical map CM, \cite{DuttaDufty} (not
shown in figure) also gives results close to those of RPIMC, and
both classical maps have the advantage of preserving the positivity
of $g(r)$. Again, there is a significant $t$ dependence between
$t=1$ and $8$ in the range $r<1 $ for both $r_{s}=1$ and $4$. The
dielectric methods all have non-physical
negative values at short distances for larger $r_{s}$ as can be seen in the $%
r_{s}=4$ panels. STLS is least negative, though VSa is much closer
to STLS than it is to RPA.

\subsection{Compressibility}

Our VSa, by construction, is the only approximate dielectric
function considered here that satisfies the CSR, STLS for example
does not. For comparison of VSa with the non-dielectric methods we
consider the
compressibility as calculated from the EOS as given in Eq. \ref%
{eq:kapparatio} for all methods. We evaluate the required
derivatives for the compressibility from the $f_{xc}$ fits mentioned
above. CHNC is not shown as the $f_{xc}$ fit has some irregularity
that show up in the derivatives as occasional wiggles in the
compressibility ratio. We note that those fits were constructed for
the free energy, not the compressibility.

\begin{figure}[tbp]
\includegraphics[angle=-90,width=1.65in]{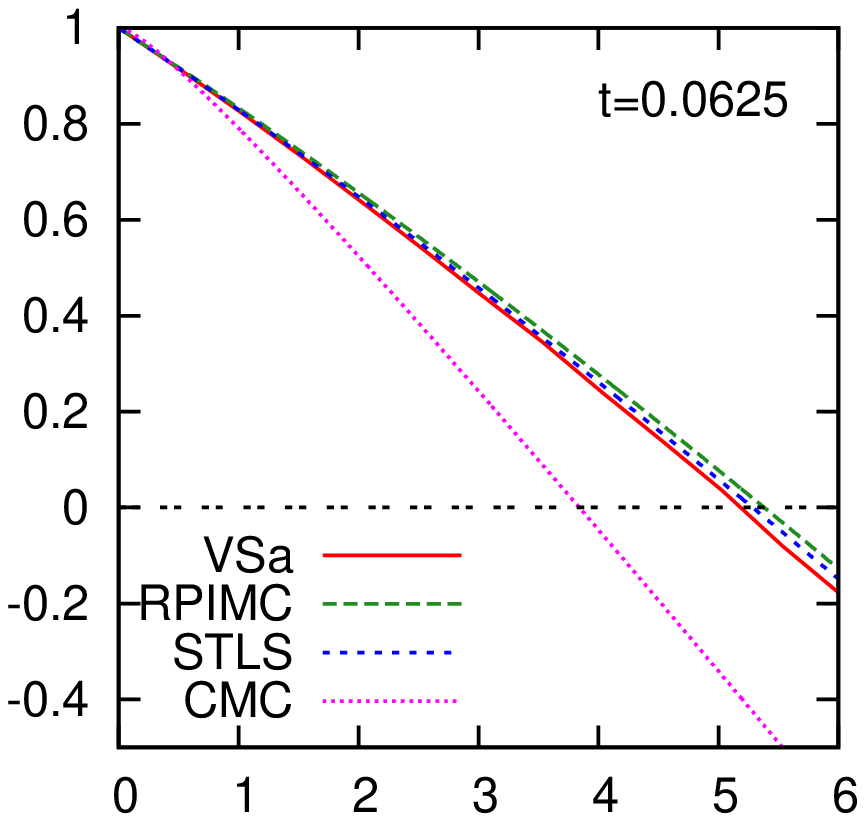} %
\includegraphics[angle=-90,width=1.65in]{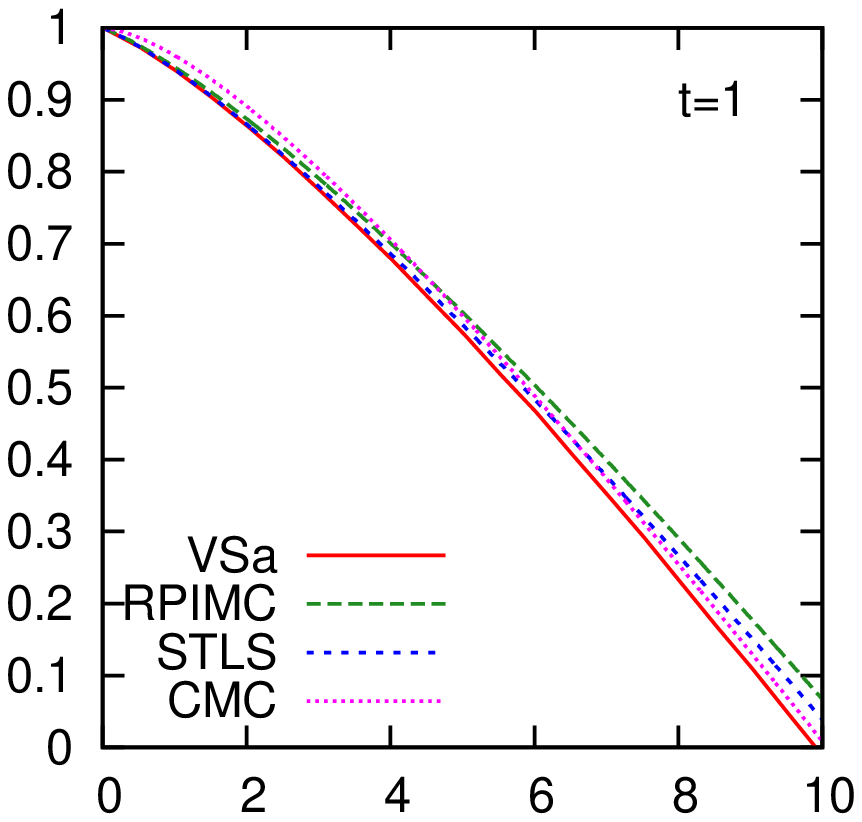} %
\includegraphics[angle=-90,width=1.65in]{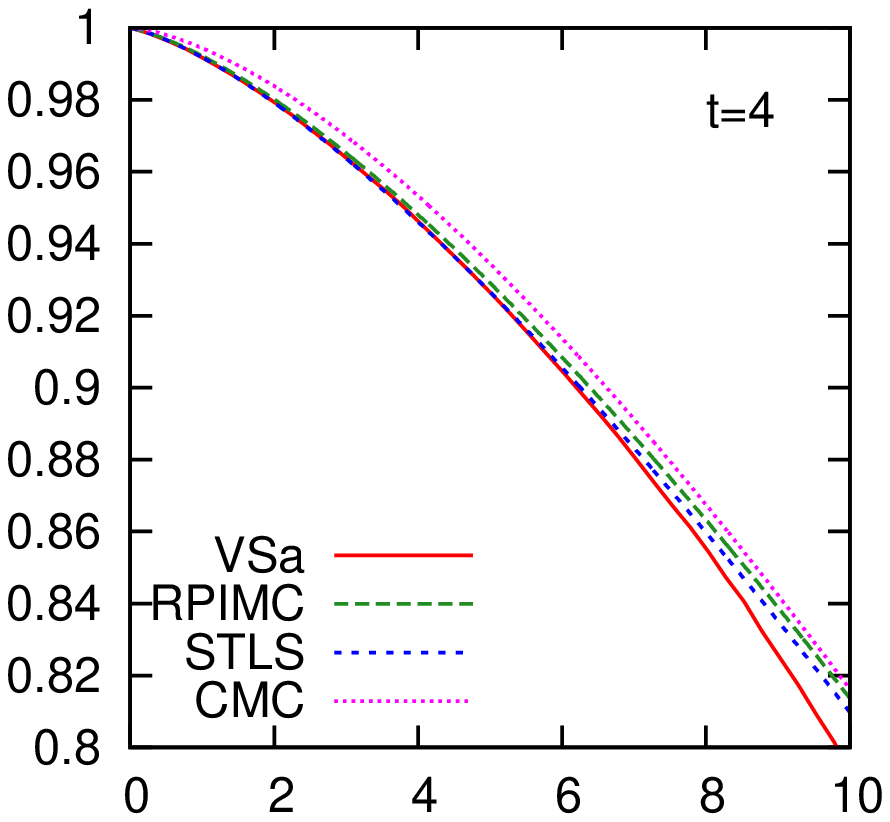} %
\includegraphics[angle=-90,width=1.65in]{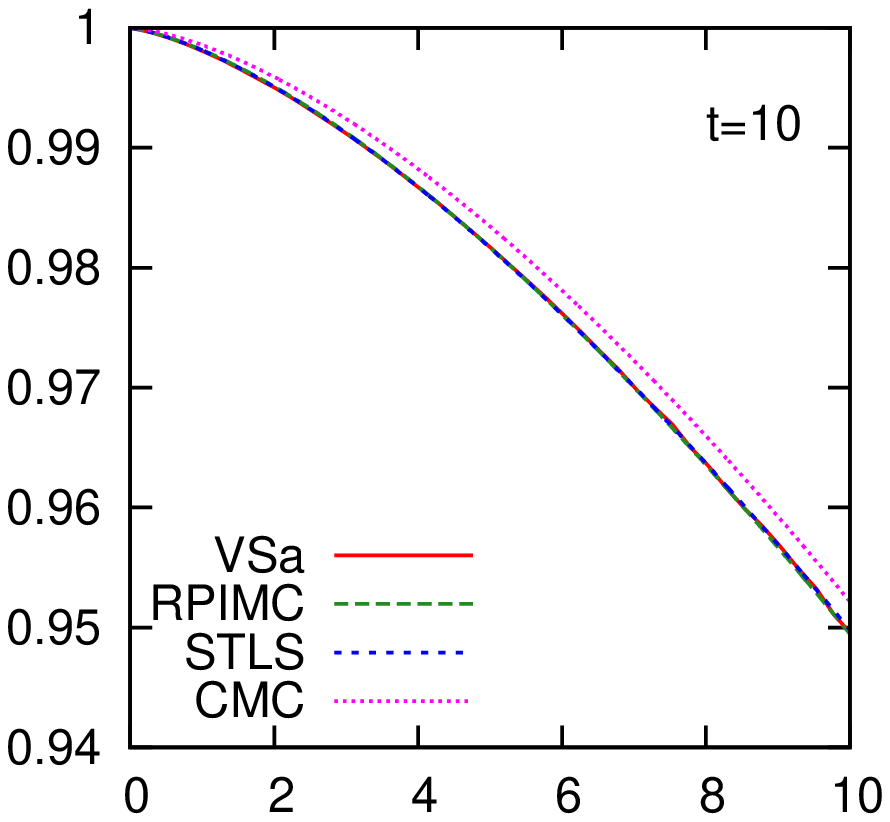}
\caption{Compressibility ratio $\protect\kappa_0/\protect\kappa$ ($y$-axis )
as a function of $r_s$ ($x$-axis) for given $t$.}
\label{fig:comp1}
\end{figure}

In Fig. \ref{fig:comp1} the compressibility ratio $\kappa
_{0}/\kappa $ is plotted at several $t$ as a function of $r_{s}$. A
first surprising observation is that the purely classical simulation
(CMC) provides semi-quantitative agreement with the quantum theories
and simulations, except at the smallest $t$ shown. At the lowest
temperature all of the quantum methods are close to the original VS
$T=0$ results, crossing zero just above $r_{s}=5$. As with the
$f_{xc}$ shown in Fig. \ref{fig:fxc1}, the STLS results lie in
between the VSa and the RPIMC results. At the highest temperature
these three results are essentially indistinguishable.

\begin{table}[b]
\caption{Values of $r_s$ for which the compressibility becomes negative for
several $t$. Also shown are Coulomb coupling constant and Debye-H\"uckel
parameter evaluated at the VSa $r_s$.}
\label{table:1}
\begin{ruledtabular}
\begin{tabular}{c|cccccc}
   t & VSa & STLS &  RPIMC & CMC & $\Gamma$ & $\lambda_D$ \\
  \hline
  0.0625 & 5.23  & 5.29  & 5.38  & 3.85  & 45.4 & 2.23 \\
  1      & 9.88  & 10.3  & 10.6  & 10.1  & 5.36 & 0.406 \\
  4      & 33.2  & 35.0  & 35.2  & 34.4  & 4.51 & 0.111 \\
  10     & 82.8  & 86.0  & 85.4  & 84.5  & 4.49 & 0.044 \\

\end{tabular}
\end{ruledtabular}
\end{table}

An interesting feature of the UEG is that at all temperatures there
is a maximum $r_{s}$ beyond which the compressibility becomes
negative, signaling an instability of the UEG system. At zero
temperature this maximum is just above $r_{s}=5$. For the case of
real metals, which of course are not true UEGs, Cs has the largest value at $%
r_{s}=5.63$ \cite{Mahan}. This instability is far below the density
for the onset of Wigner crystallization. In Table \ref{table:1} we
record this maximum $r_{s}$ as given by VSa, STLS, RPIMC, and CMC
(which includes classical strong coupling contributions beyond DH
\cite{Hansen}). Also shown are the Coulomb coupling constant,
$\Gamma=0.543\,r_{s}/t $ and the Debye-H\"{u}ckel parameter,
$\lambda _{D}=1.276/\sqrt{tr_{s}}$, both evaluated at the value of
$r_{s}$ for the instability predicted from VSa.

\subsection{Classical Limit}

\begin{figure}
\includegraphics[angle=-90,width=3.25in]{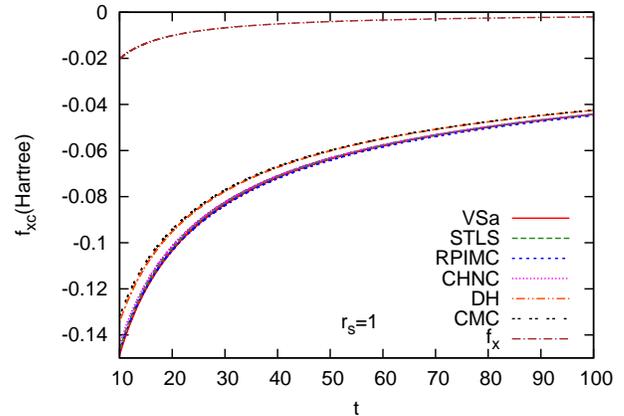} %
\caption{Comparison for large $t$ at $r_s=1$ for the XC free energy.}
\label{fig:limt}
\end{figure}

The ideal Fermi gas thermodynamics depends on $n$ and $T$ only through $t$,
and at $t=10$ the classical limit is approached. For the interacting UEG,
properties depend on both $t$ and $r_{s}$ through the Coulomb interactions
and the large $t$ classical limit is not uniform in $r_{s}$. For fixed
$r_{s}$ there is a sufficiently large $t$ above which the classical limit applies.
However, within this limit the DH limit need not apply. The latter
requires in addition small $\Gamma$. In order to examine the classical
limit we consider the case $r_{s}=1$ in the large $t$ limit. In this limit
correct results should come into agreement with the Debye-H\"{u}ckel result
since $\Gamma $ is small, and the fits are mostly constructed to do so.
Figure \ref{fig:limt} shows the XC free energy for
$r_{s}=1$ and $t$ from $10-100$. The XC free energy shows agreement between
all of the quantum methods VSa, STLS, RPIMC, and CHNC. Additionally
the classical DH and CMC are in very good agreement with each other, with small differences
becoming visible below $t=25$. The difference between those
classical results and the quantum results is mainly due to the exchange contribution,
$f_{x}$, which is shown near the top of the plot.

\section{Conclusions}

In this work we have presented calculations for the uniform electron gas
from an approximate dielectric function method based on a finite temperature
version of the Vashista-Singwi static local field correction, modified to
enforce the compressibility sum rule at all $t,r_{s}$. This sum rule is
violated by the orginal $t=0$ VS(0), and by previous finite temperature RPA
and STLS approximations.

We have made comparisons of equilibrium structure and thermodynamics
calculations with other finite temperature RPA, STLS, and classical
mapping methods, and with restricted path integral Monte Carlo
results. Our VSa method in general produces results between RPA and
STLS, though closer to STLS, for the UEG properties considered
$G(k)$, $S(k)$, $g(r)$, $f_{xc}$, and $\kappa $. For $f_{xc}$
dielectric methods and classical map methods are similar, as is our
fit for RPIMC.  The results
for the compressibility follow these same trends. This includes some
deviation of VSa from STLS and RPIMC methods for intermediate $r_s$,
and $t$ values. For $g(r)$ the dielectric methods produce unphysical
negative values at small $r$ and large $r_{s}$, while both the RPIMC
and classical map methods produce non-negative $g(r)$.

Finally we see that STLS and RPIMC in fact cross validate each other
very nicely. It has long been known that STLS gives quite good zero
$T$ XC energies compared to QMC results, and this seems to be true
for finite temperatures as well. This good agreement is also seen to
apply for the interaction energy (e.g. Fig. \ref{fig:eint}). This
contrasts somewhat with the comparisons of XC energy and RPIMC in
the recent fit analysis of Brown et. al \cite{Brownetal}. Perhaps
surprisingly, the VSa with internal consistency for the
compressibility sum rule deviates somewhat more from the RPIMC
results than its underlying STLS method without this consistency.
The simplest dielectric approach, RPA, is not shown here as the
deviations from other methods is generally quite large.

In summary we have compared the most accurate approximations of
$f_{xc}$ and found them close, but in particular STLS and RPIMC seem
to pin down the correct results. This lends theoretical support for
the simulations and their extension by the fit for the RPIMC
$f_{xc}$ given here. An important application, to be discussed
further elsewhere, is the implementation as a local density functional,
and construction of more complex functionals 
needed for finite temperature DFT.

\section{Acknowledgements}

This research has been supported by US DOE Grant DE-SC0002139. T.S.
acknowledges support by the NNSA of the US DOE at Los Alamos
National Laboratory under Contract No. DE-AC52-06NA25396. The
authors thank S. Dutta for providing CHNC calculations of the pair
correlation function.

\appendix*

\section{Fits for the exchange correlation free energy}

\begin{table}
\caption{Fit parameters for the exchange-correlation free energy for STLS, VSa, and RPIMC given by the Eqs. \ref{eq:eintfit}-\ref{eq:fxcfit}. STLS parameters as given in Ref. \onlinecite{Ichimaru93}.}
\label{table:param}
\begin{ruledtabular}
\begin{tabular}{clll}
        & STLS & VSa & RPIMC \\
        \hline
   $x_1$  & 3.4130800$\times 10^{-1}$ & 1.8871493$\times 10^{-1}$ & 3.4130800$\times 10^{-1}$ \\
   $x_2$  & 1.2070873$\times 10^{1}$ & 1.0684788$\times 10^{1}$ & 8.7719094$\times 10^{1}$ \\
   $x_3$  & 1.148889$\times 10^{0}$ & 1.1088191$\times 10^{2}$ & 4.4699486$\times 10^{3}$ \\
   $x_4$  & 1.0495346$\times 10^{1}$ & 1.8015380$\times 10^{1}$ & 3.4072692$\times 10^{2}$ \\
   $x_5$  & 1.326623$\times 10^{0}$ & 1.2803540$\times 10^{2}$ & 5.1614521$\times 10^{3}$ \\
   $x_6$  & 8.72496$\times 10^{-1}$ & 8.3331352$\times 10^{-1}$ & 8.6415253$\times 10^{-1}$ \\
   $x_7$  & 2.5248$\times 10^{-2}$ & -1.1179213$\times 10^{-1}$ & -9.2236194$\times 10^{-2}$ \\
   $x_8$  & 6.14925$\times 10^{-1}$ & 6.1492503$\times 10^{-1}$ & 6.1492503$\times 10^{-1}$ \\
   $x_9$  & 1.6996055$\times 10^{1}$ & 1.6428929$\times 10^{1}$ & 2.5191969$\times 10^{1}$ \\
   $x_{10}$ & 1.489056$\times 10^{0}$ & 2.5963096$\times 10^{1}$ & 1.8208366$\times 10^{1}$ \\
   $x_{11}$ & 1.010935$\times 10^{1}$ & 1.0905162$\times 10^{1}$ & 1.8659964$\times 10^{1}$ \\
   $x_{12}$ & 1.22184$\times 10^{0}$ & 2.9942171$\times 10^{1}$ & 1.8463421$\times 10^{1}$ \\
   $x_{13}$ & 5.39409$\times 10^{-1}$ & 5.3940898$\times 10^{-1}$ & 5.3940898$\times 10^{-1}$ \\
   $x_{14}$ & 2.522206$\times 10^{0}$ & 5.8869626$\times 10^{4}$ & 2.9390225$\times 10^{2}$ \\
   $x_{15}$ & 1.78484$\times 10^{-1}$ & 3.1165052$\times 10^{3}$ & 1.1501733$\times 10^{1}$ \\
   $x_{16}$ & 2.555501$\times 10^{0}$ & 3.8887108$\times 10^{4}$ & 3.2847098$\times 10^{2}$ \\
   $x_{17}$ & 1.46319$\times 10^{-1}$ & 2.1774472$\times 10^{3}$ & 8.7963510$\times 10^{0}$ \\
\end{tabular}
\end{ruledtabular}
\end{table}

An effective fitting procedure for STLS calculations has been given
by Ichimaru in Ref. \onlinecite{Ichimaru93} page 290; that fit has
been used for all STLS plots above. We extend that method in this
Appendix to the VSa calculations and RPIMC results. First, the same
functional form is chosen for the interaction energy, expressed in
terms of $\Gamma, t$ instead of $r_s,t$, and a least squares fitting
for the parameters is performmed.  With the coefficients known and 
dependence on $\Gamma$ displayed explicitly, the coupling
constant integration of Eq. \ref{eq:fxc1} can be performed to get
the exchange correlation free energy per particle, $f_{xc}$.

The interaction energy per particle is given in Hartree units by
\begin{align}
  e_{\rm int}(r_s,t) = -\frac{\Gamma}{\beta} \frac{a(t)+b(t)\sqrt{\Gamma}+c(t)\Gamma}
  {1+d(t)\sqrt{\Gamma}+e(t)\Gamma}
  \label{eq:eintfit}
\end{align}
Here $a(t)$ is given by the exchange parametrization given in Ref.
\onlinecite{PDw84} as
\begin{align}
  a(t)=&0.610887 \tanh{ \left( \frac{1}{t} \right)  } \times \nonumber \\
  &\frac{0.75 + 3.04363 t^2-0.09227 t^3+1.7035 t^4}
  {1 + 8.31051 t^2 + 5.1105 t^4} \;.
\end{align}
Terms $b$-$e$ are given by
\begin{align}
  b(t)=&\sqrt{t}\tanh{ \left( \frac{1}{\sqrt{t}} \right)  }
  \frac{x_1 + x_2 t^2 + x_3 t^4}{1+x_4 t^2 +x_5 t^4} \\
  c(t)=&\left[ x_6 +x_7 \exp \left( -\frac{1}{t} \right) \right] e(t) \\
  d(t)=&\sqrt{t}\tanh{ \left( \frac{1}{\sqrt{t}} \right)  }
  \frac{x_8 + x_9 t^2 + x_{10} t^4}{1+x_{11} t^2 +x_{12} t^4} \\
  e(t)=&t \tanh{ \left( \frac{1}{t} \right) }
  \frac{x_{13} + x_{14} t^2 + x_{15} t^4}{1+x_{16} t^2 +x_{17} t^4}
\end{align}
The fit parameters are chosen to give the correct high $t$ limit. In
table \ref{table:param} we provide our new fit parameters for both
VSa and RPIMC, as well as those for STLS from Ref.
\onlinecite{Ichimaru93}.  The coupling constant integration to give
the XC free energy is also given in Ref. \onlinecite{Ichimaru93} as
\begin{widetext}
\begin{align}
  f_{xc}(r_s,t)=& -\frac{c}{e}\frac{\Gamma}{\beta} -\frac{2}{e} \left( b -\frac{cd}{e} \right) \frac{\sqrt{\Gamma}}{\beta}-\frac{1}{\beta e} \left[ \left(a-\frac{c}{e}\right) -\frac{d}{e}\left(b-\frac{cd}{e} \right) \right]  \ln { \left| e \Gamma + d \sqrt{\Gamma} + 1 \right| } \nonumber \\
& + \frac{2}{\beta e \sqrt{4e-d^2}} \left[ d \left(a-\frac{c}{e}\right) + \left( 2- \frac{d^2}{e} \right) \left( b -\frac{cd}{e} \right) \right] \left[ \tan^{-1} \left( \frac{2e\sqrt{\Gamma}+d}{\sqrt{4e-d^2}} \right) - \tan^{-1} \left( \frac{d}{4e-d^2} \right) \right] \;.
\label{eq:fxcfit}
\end{align}
\end{widetext}


\begin{figure*}
  \includegraphics[angle=-90,width=3.0in]{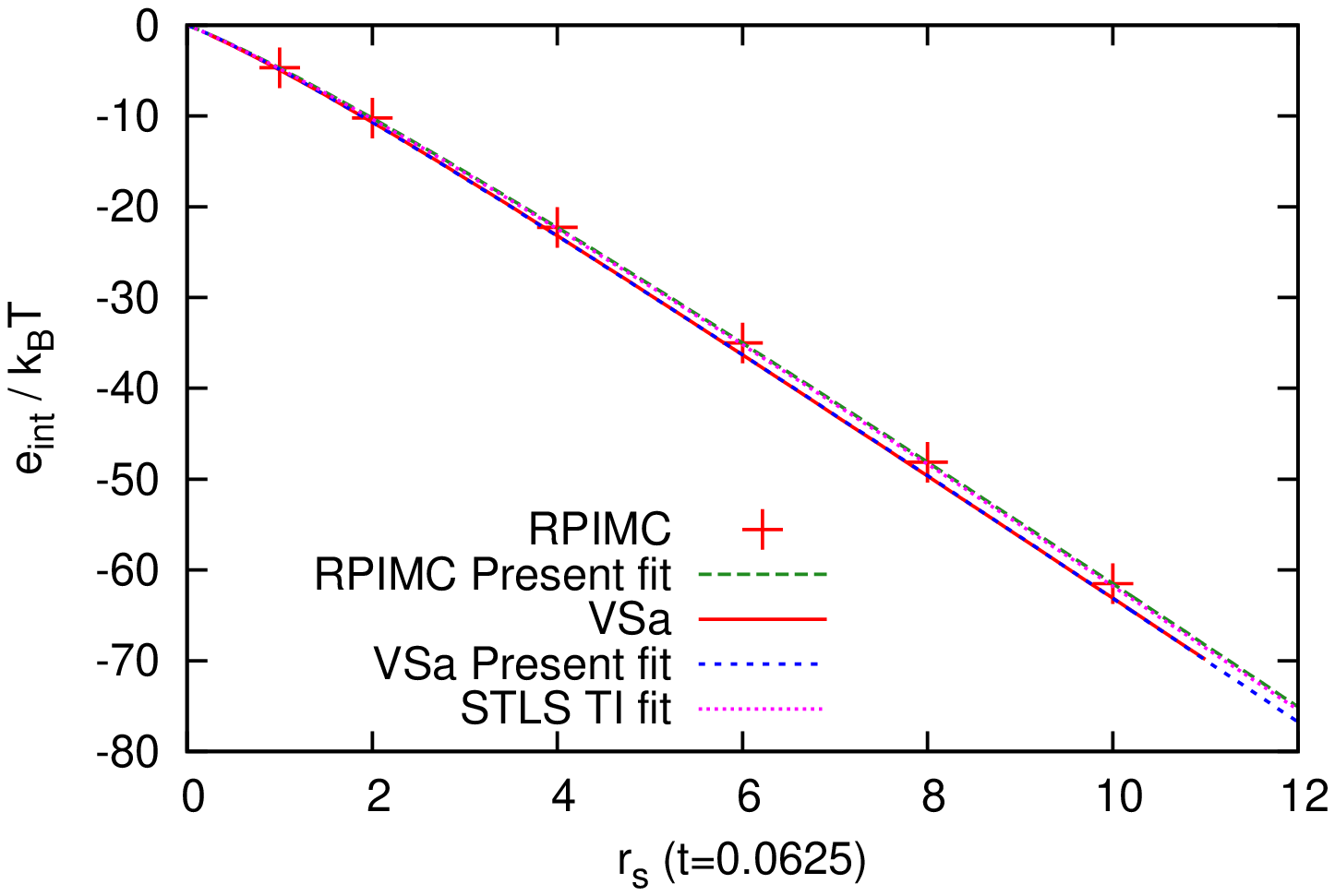}
  \includegraphics[angle=-90,width=3.0in]{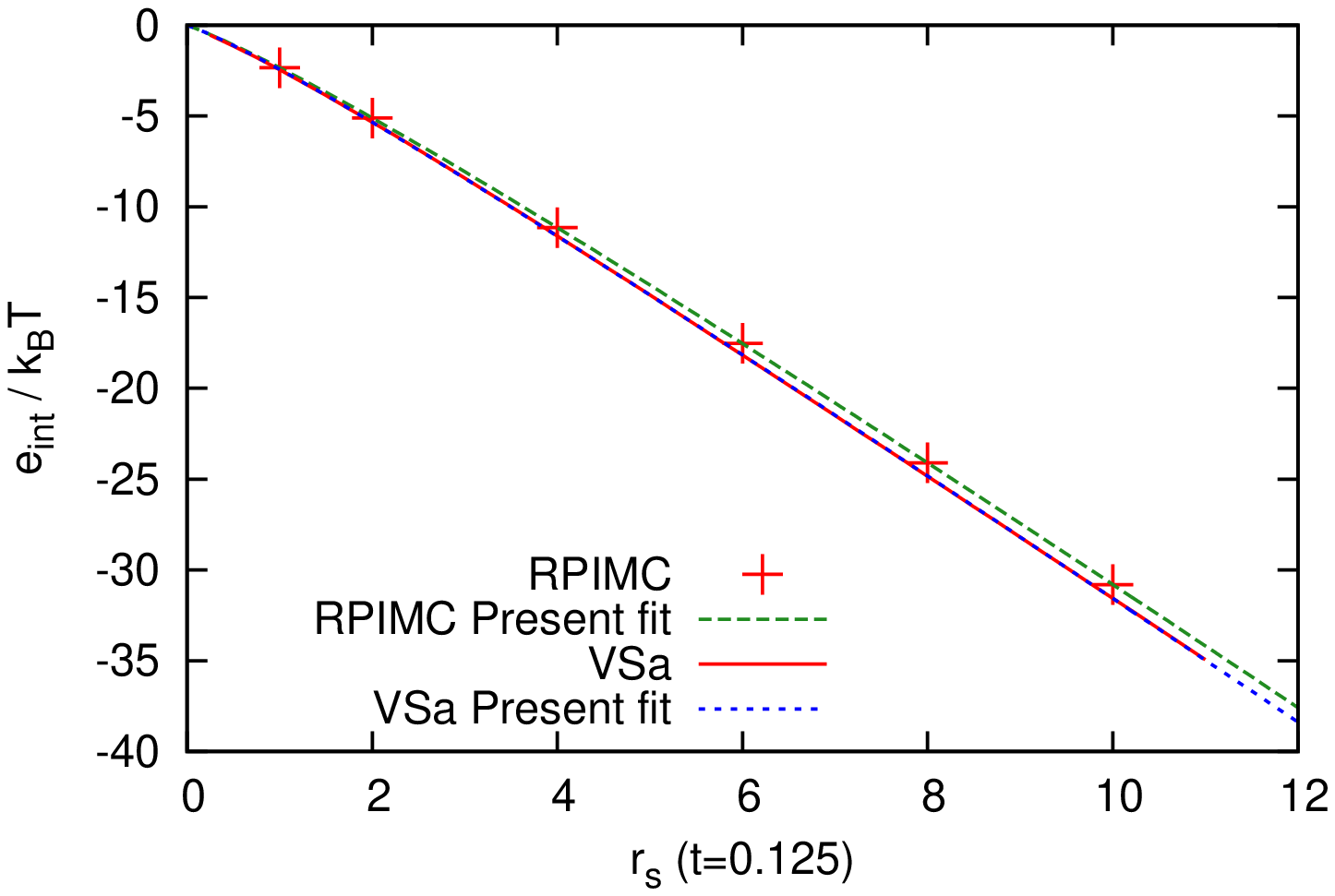}
  \includegraphics[angle=-90,width=3.0in]{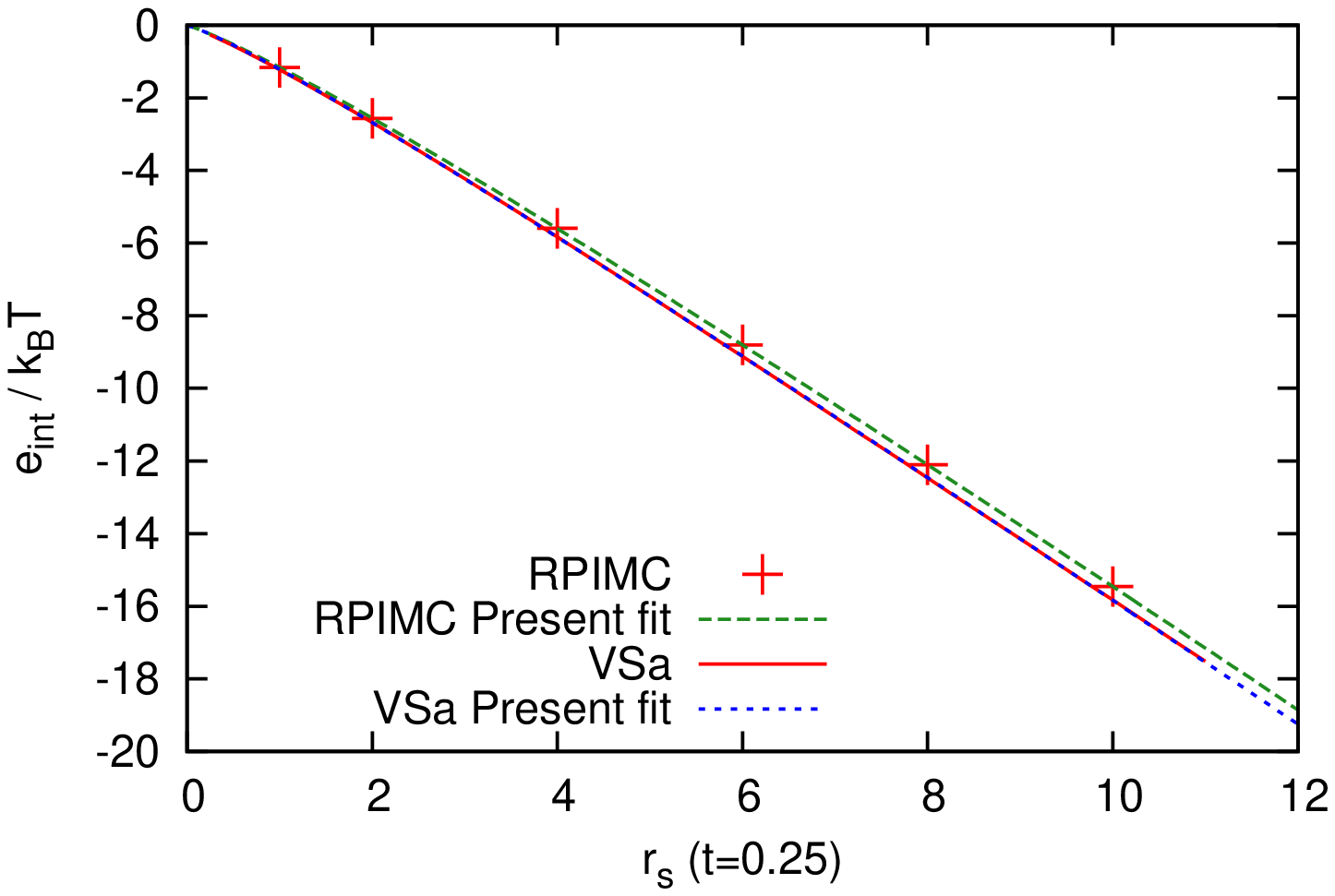}
  \includegraphics[angle=-90,width=3.0in]{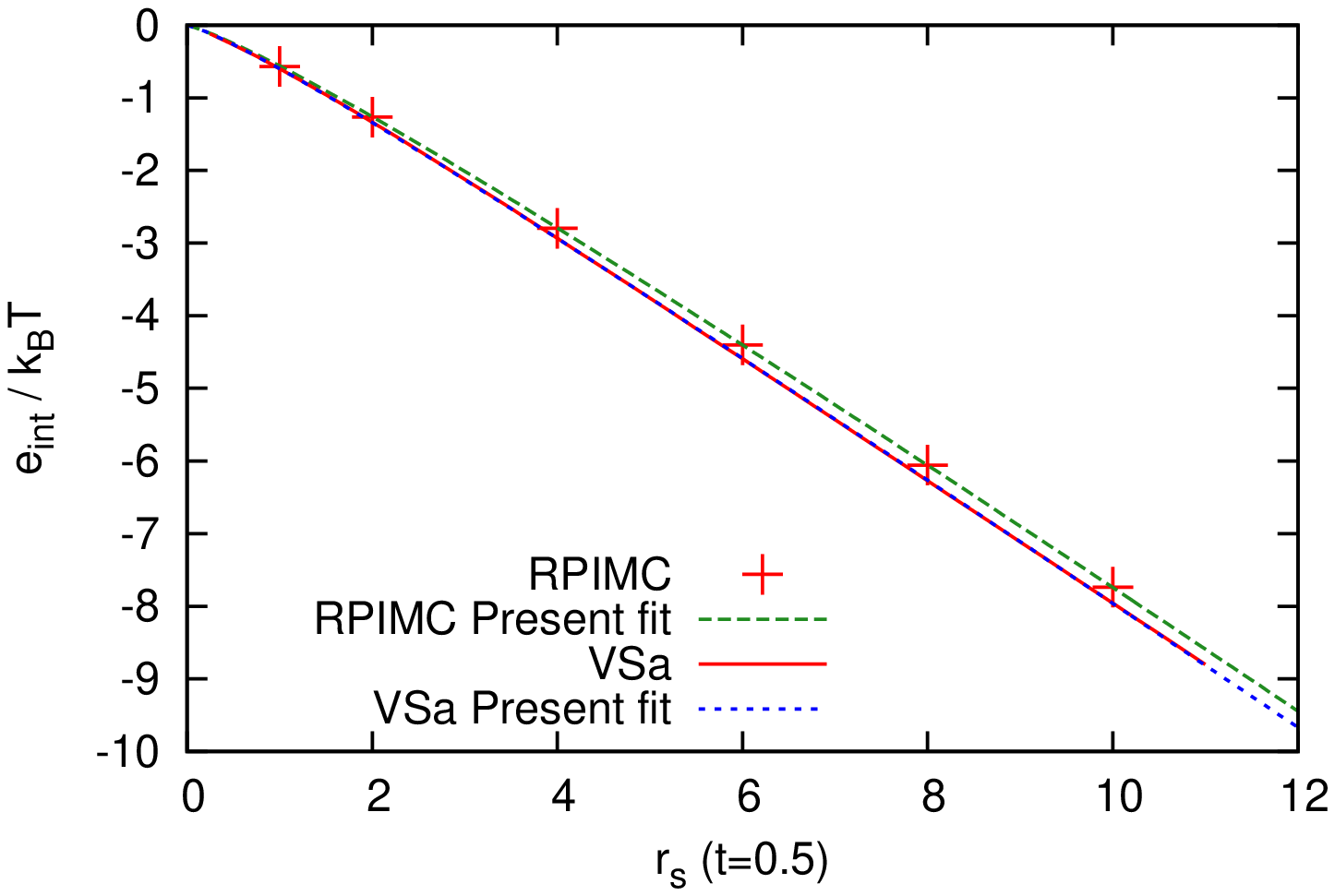}
  \includegraphics[angle=-90,width=3.0in]{eps_int-t1.0.eps}
  \includegraphics[angle=-90,width=3.0in]{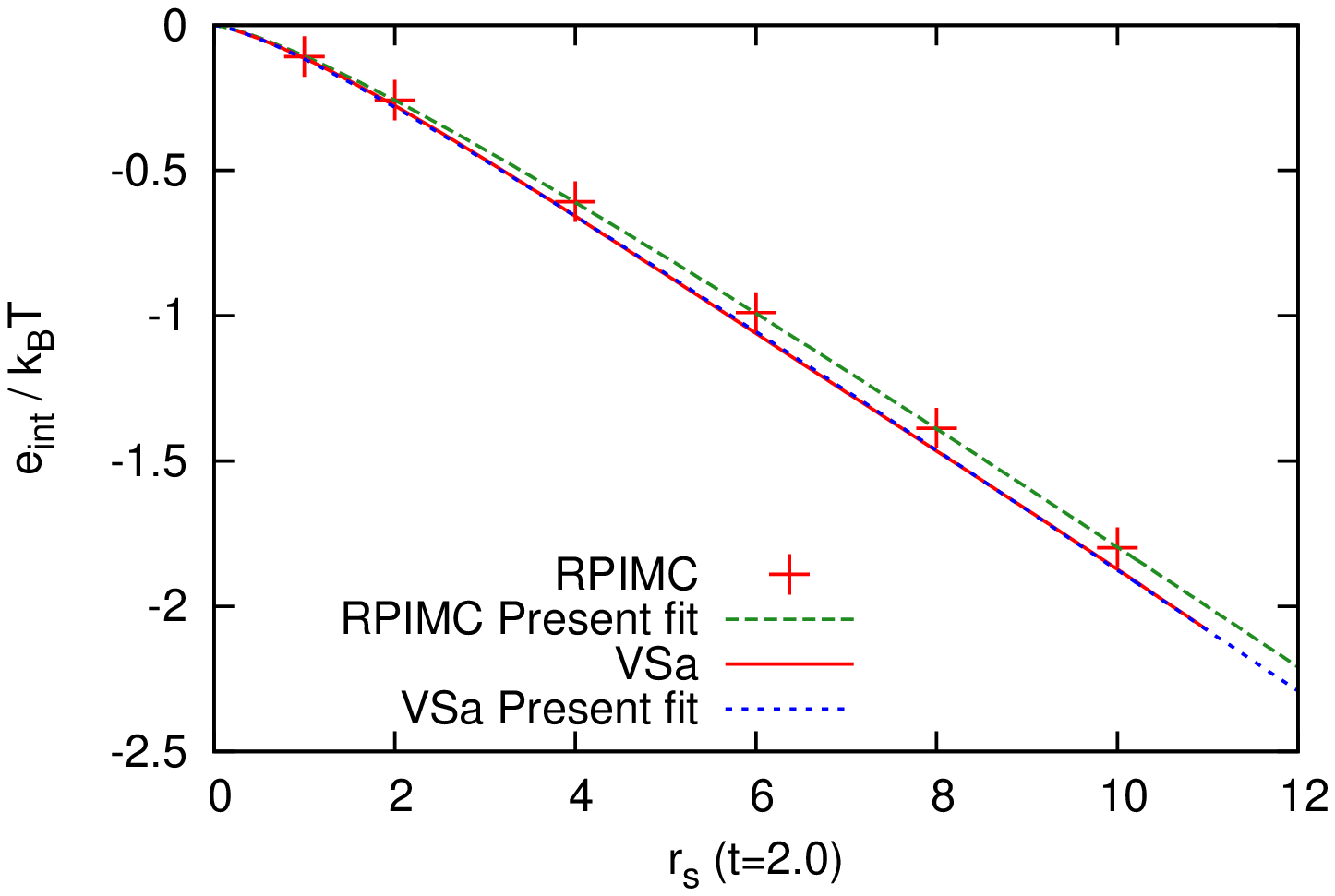}
  \includegraphics[angle=-90,width=3.0in]{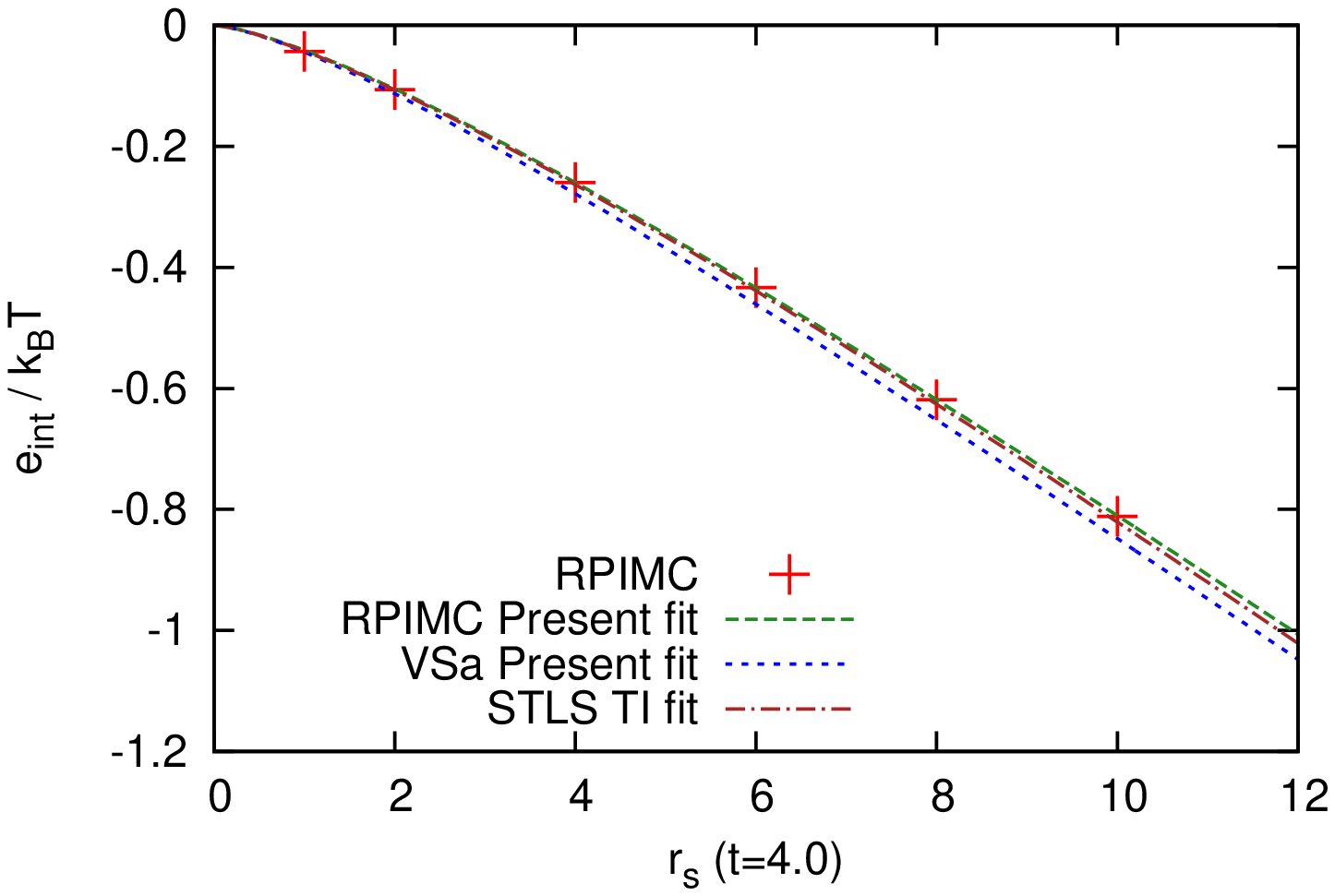}
  \includegraphics[angle=-90,width=3.0in]{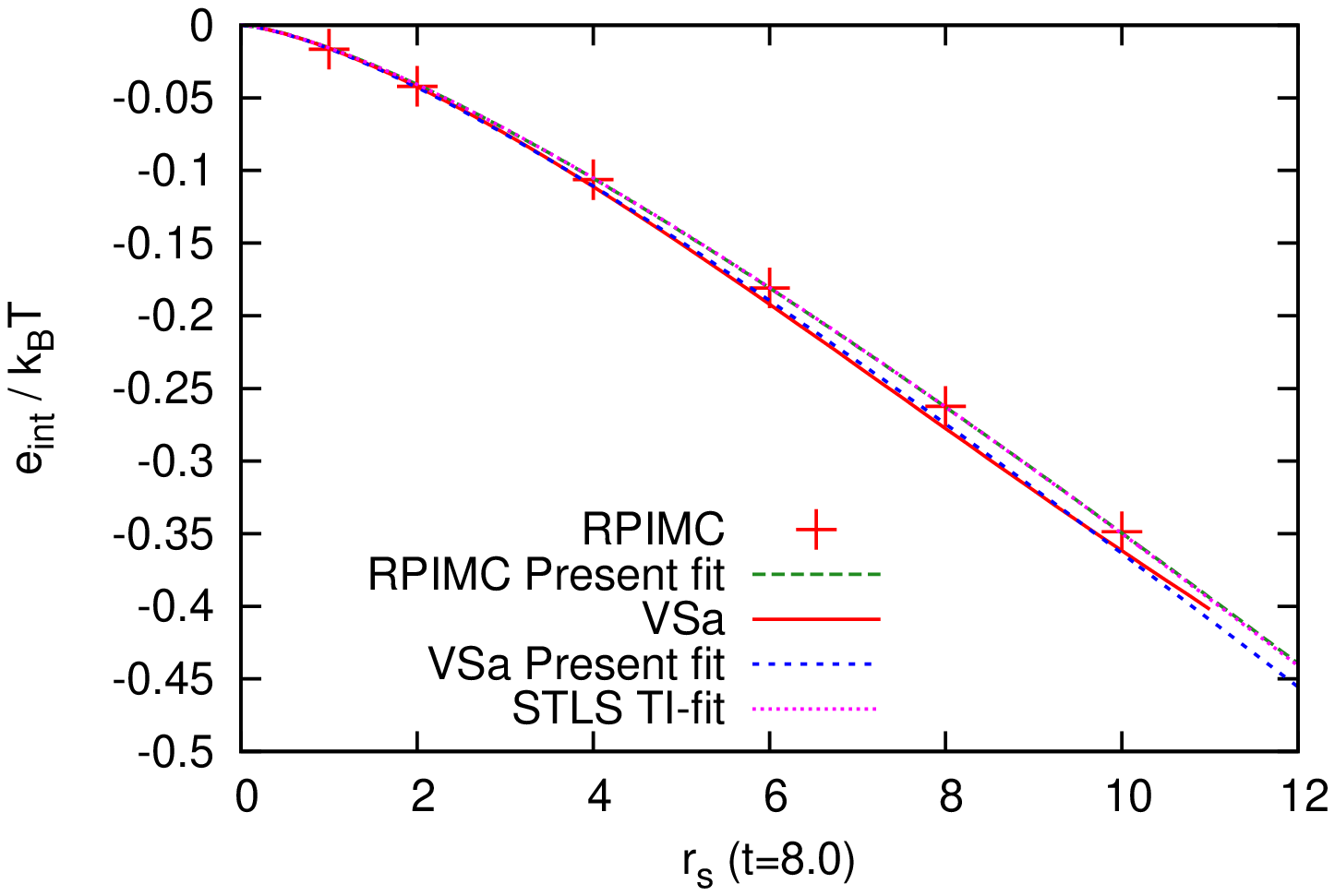}
  \caption{Interaction energy in temperature units. Raw data and fits for RPIMC and VSa, and for some STLS fits. Check of the fits is good for both RPIMC and VSa. Comparison of results shows STLS in between VSa and RPIMC, but closer to RPIMC at all $t$.}
\end{figure*}


\begin{figure*}
  \includegraphics[angle=-90,width=3.0in]{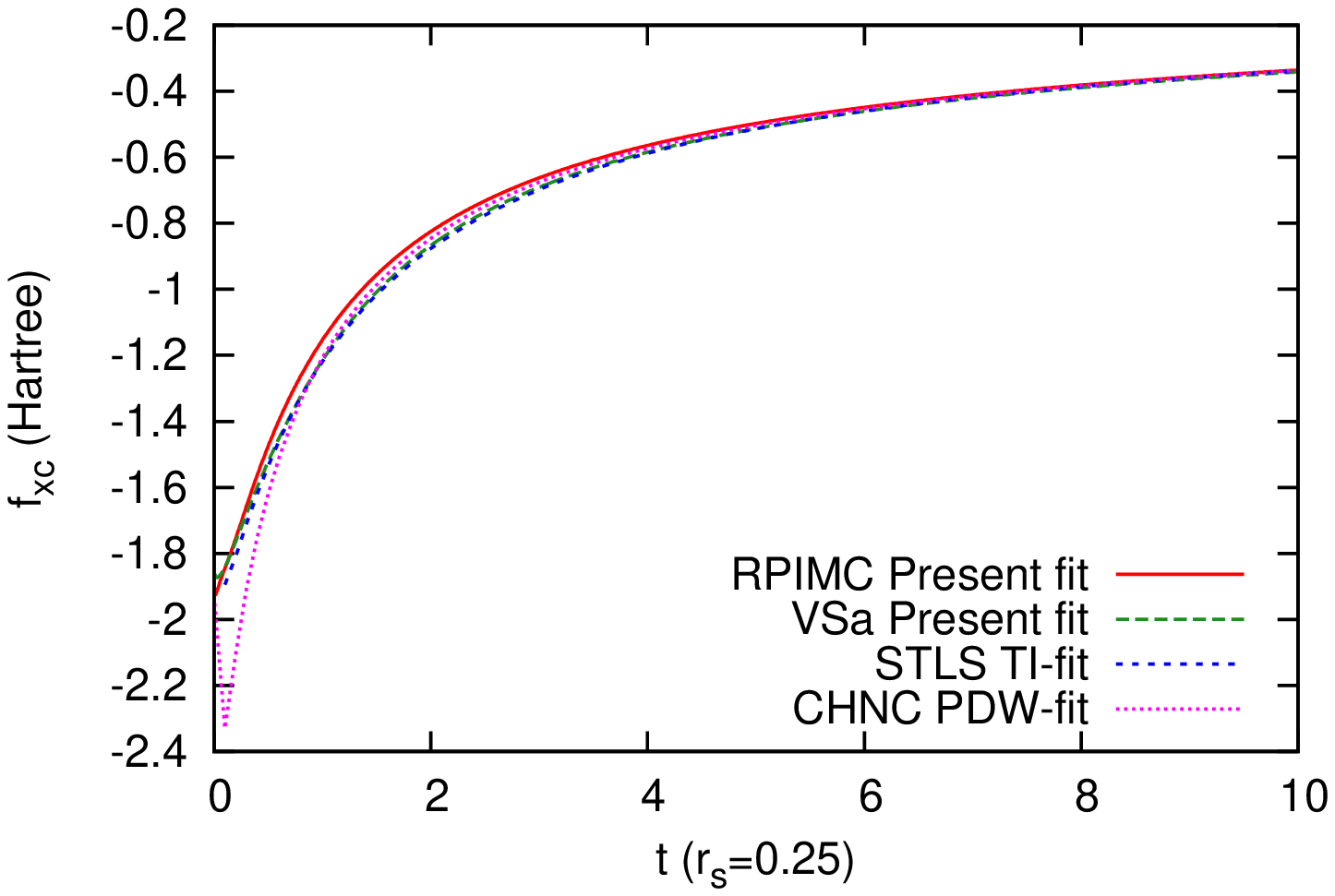}
  \includegraphics[angle=-90,width=3.0in]{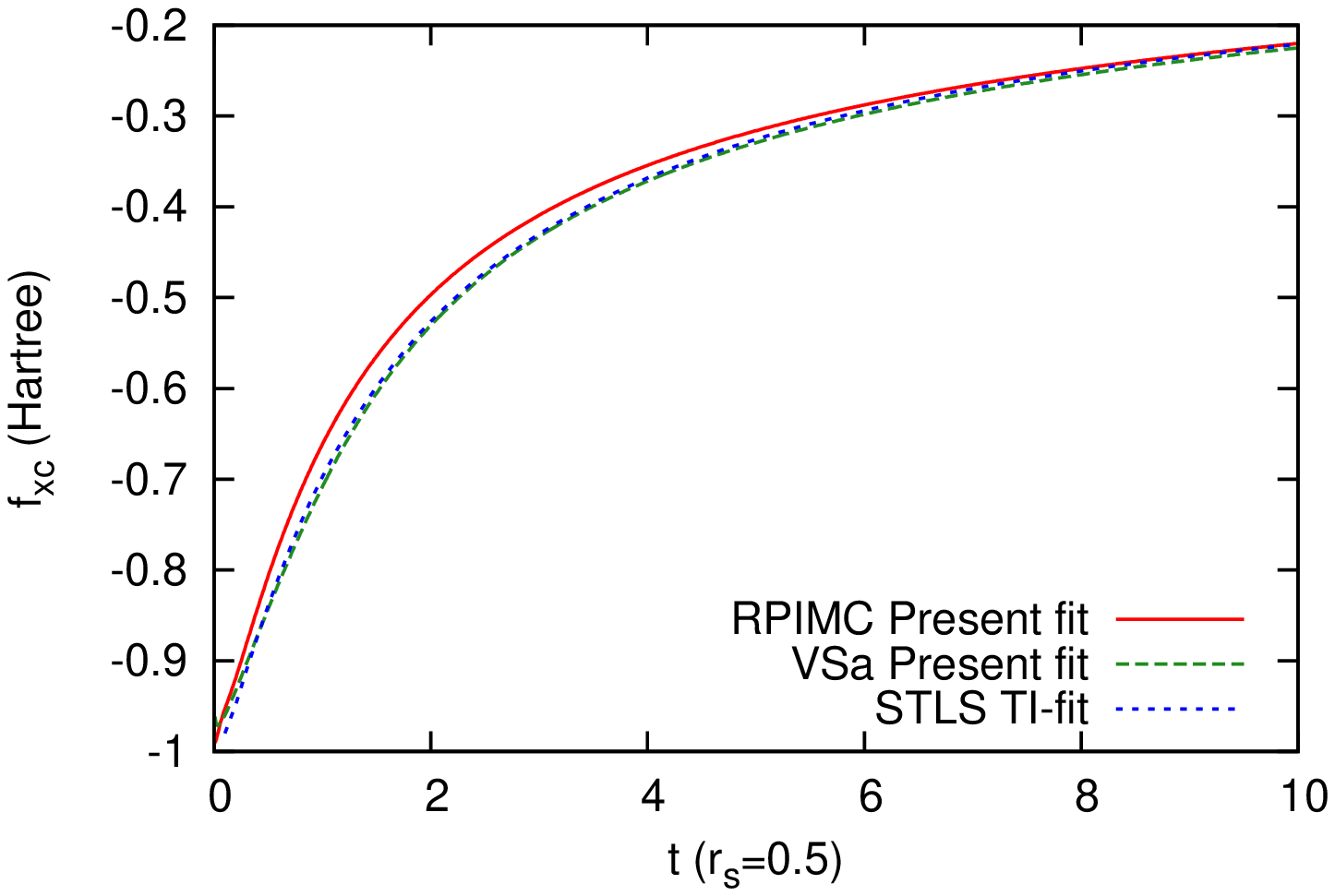}
  \includegraphics[angle=-90,width=3.0in]{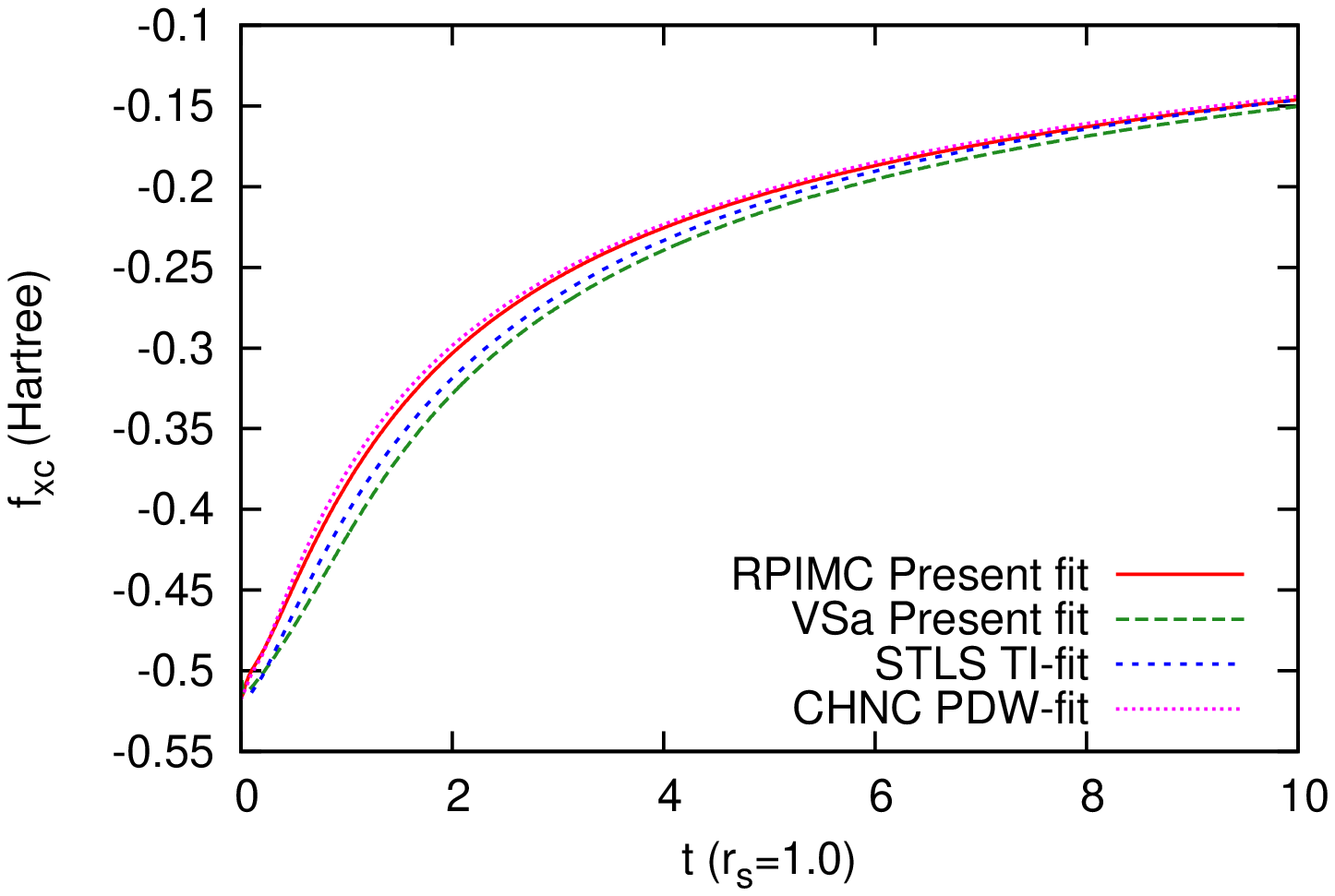}
  \includegraphics[angle=-90,width=3.0in]{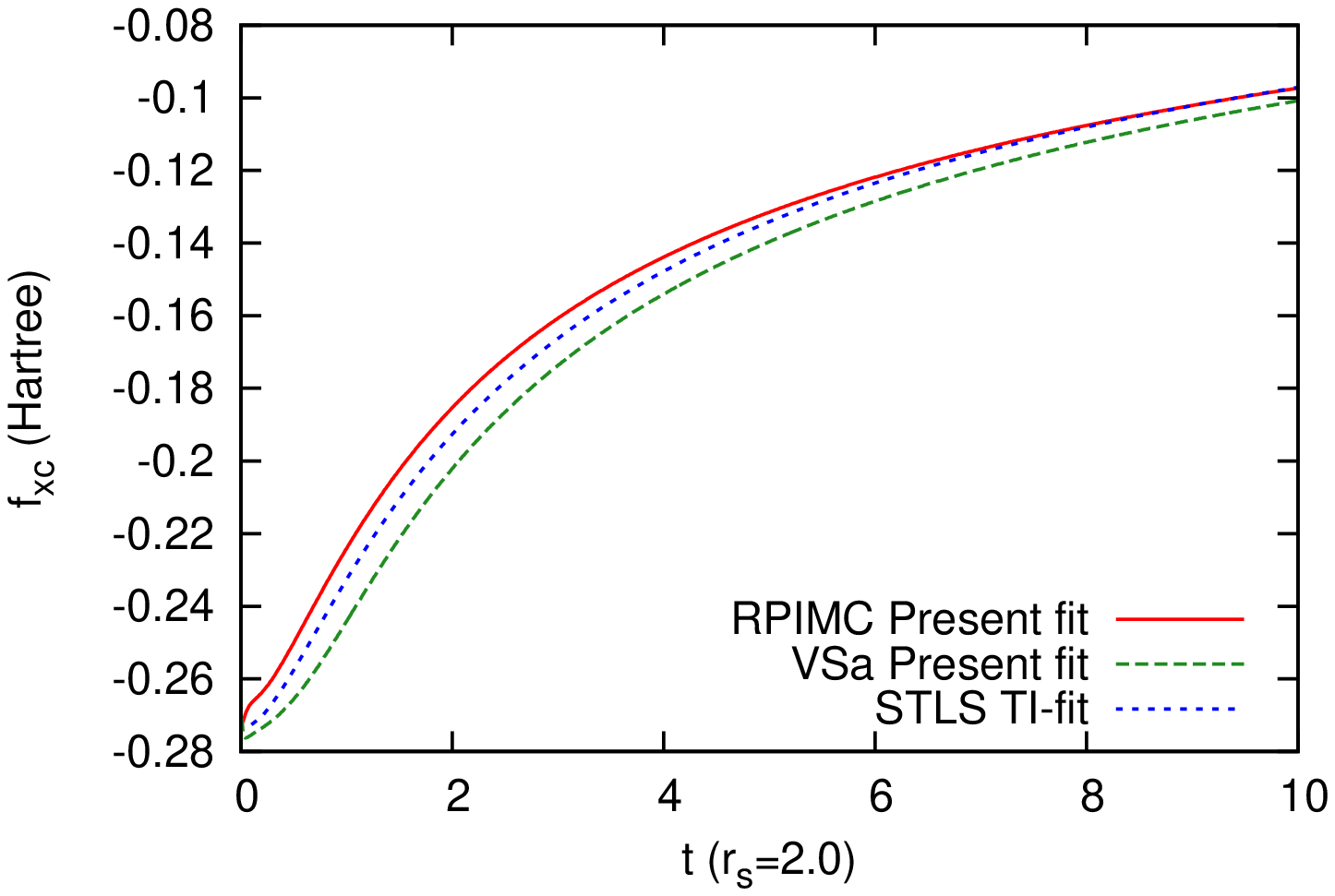}
  \includegraphics[angle=-90,width=3.0in]{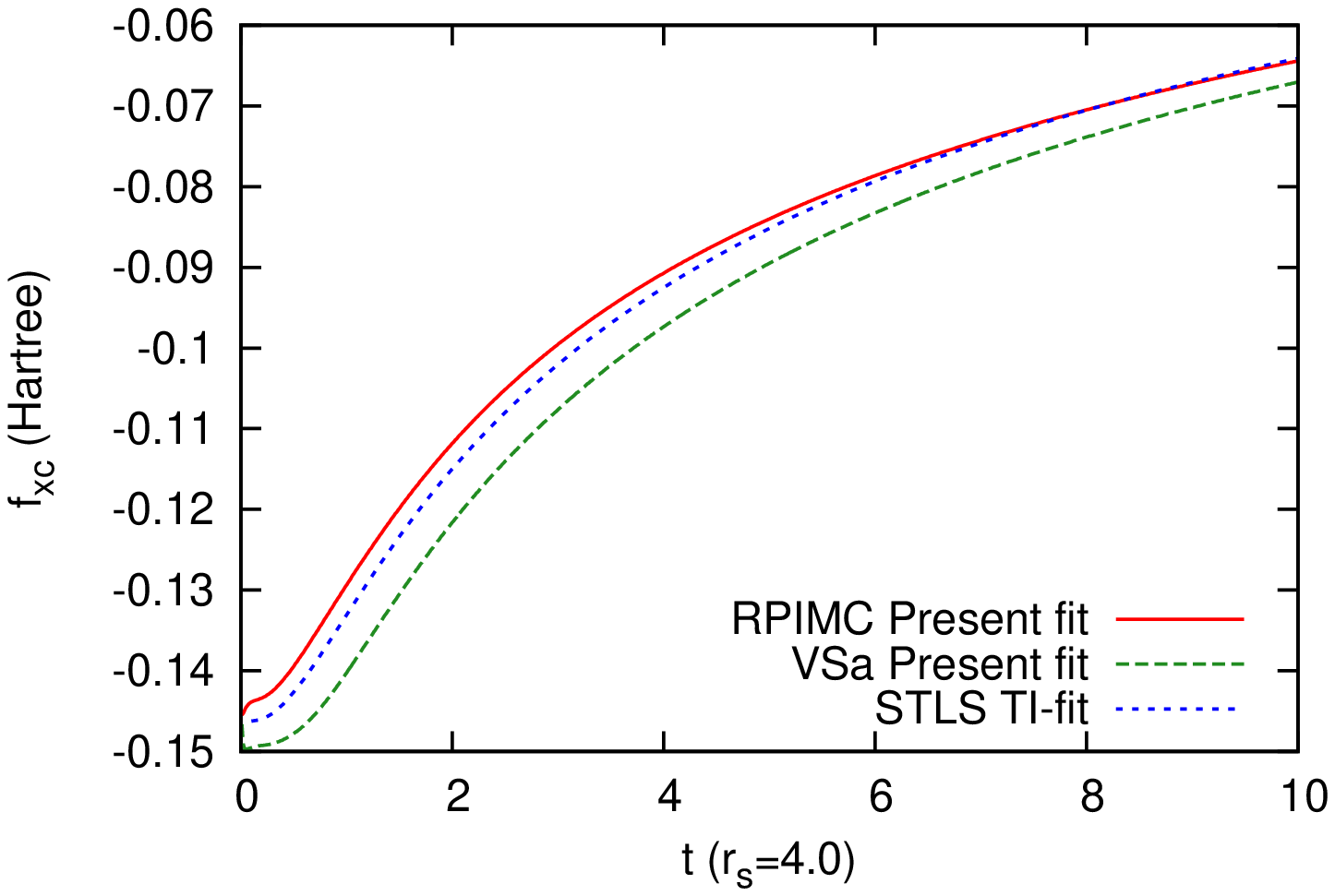}
  \includegraphics[angle=-90,width=3.0in]{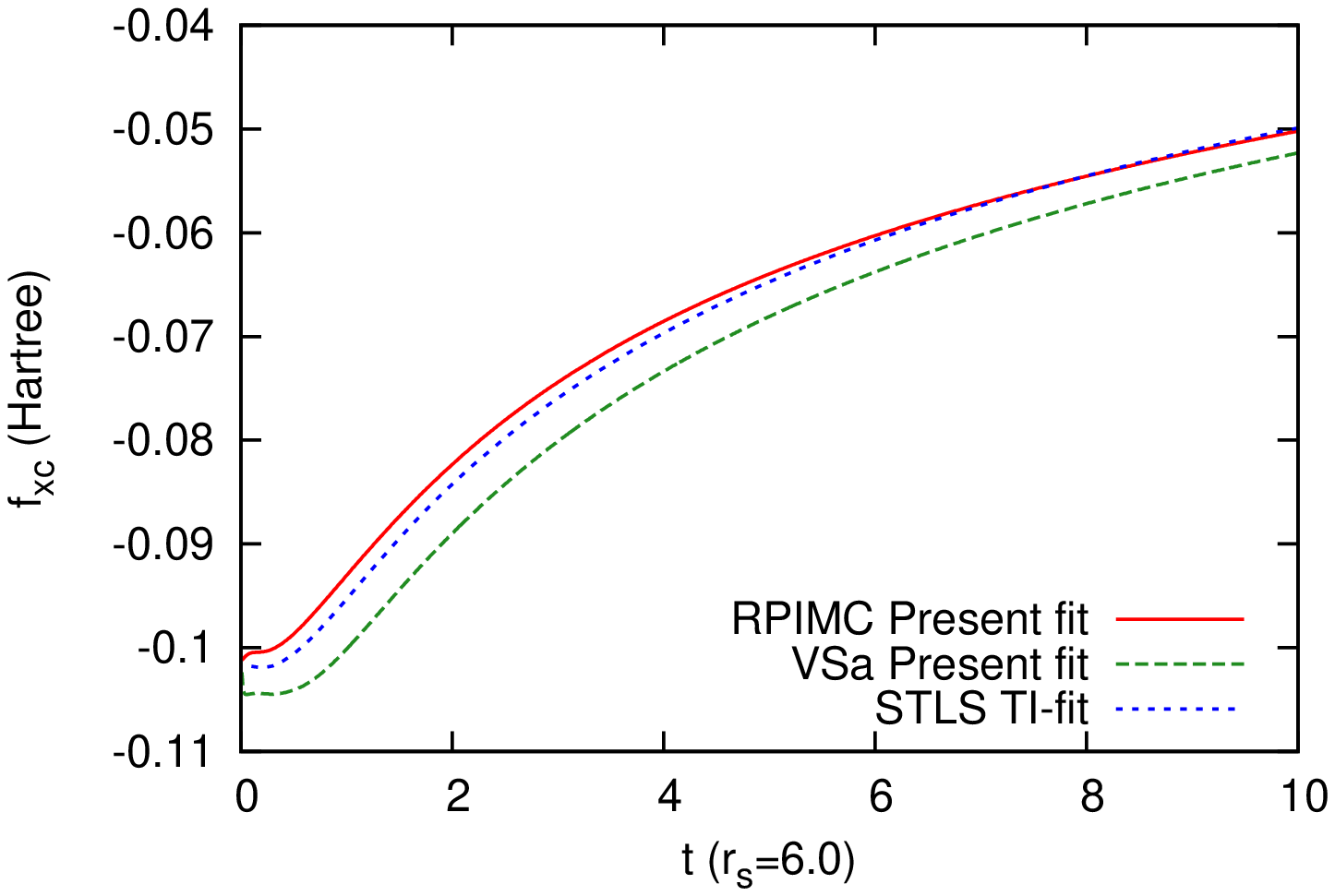}
  \includegraphics[angle=-90,width=3.0in]{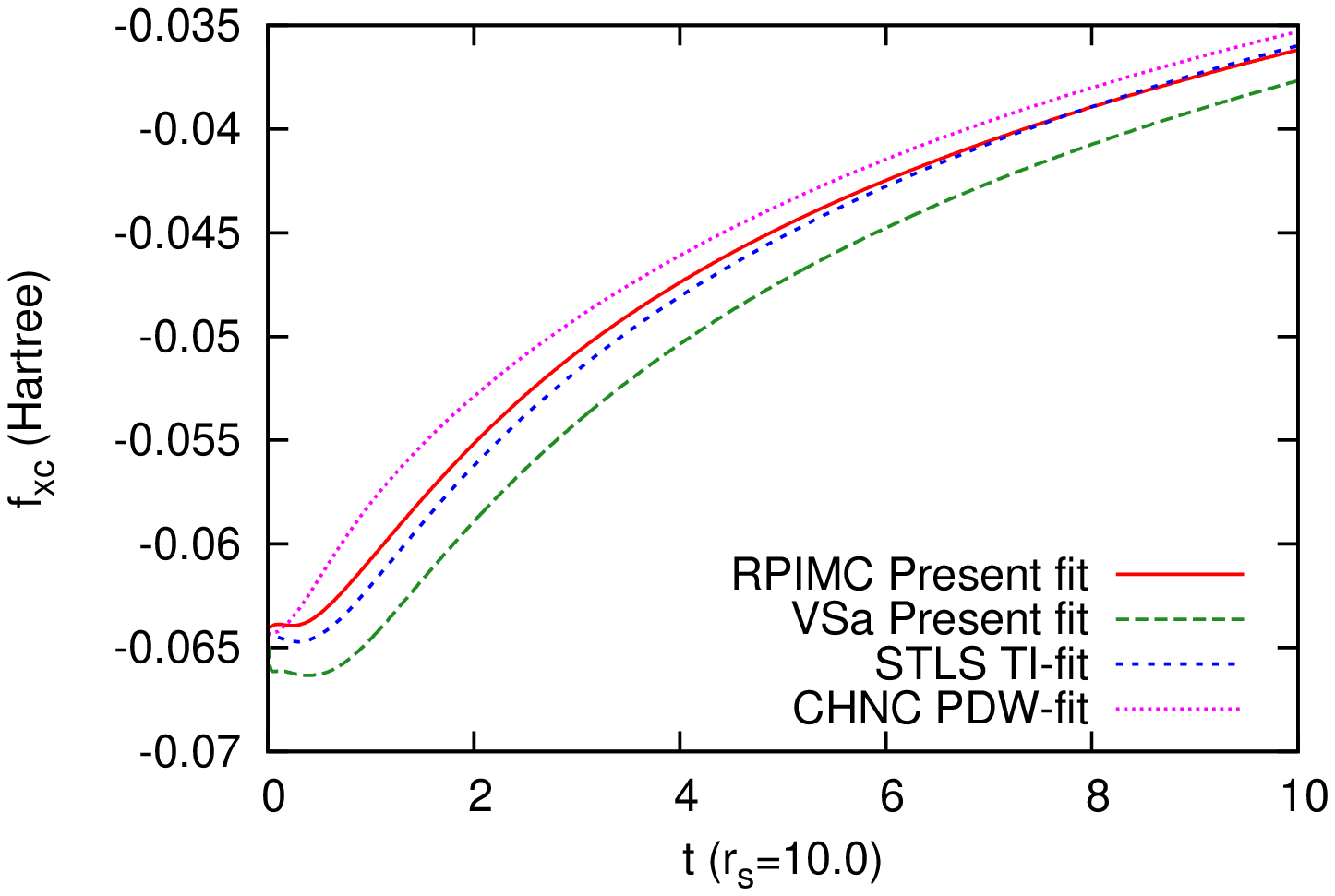}
  \includegraphics[angle=-90,width=3.0in]{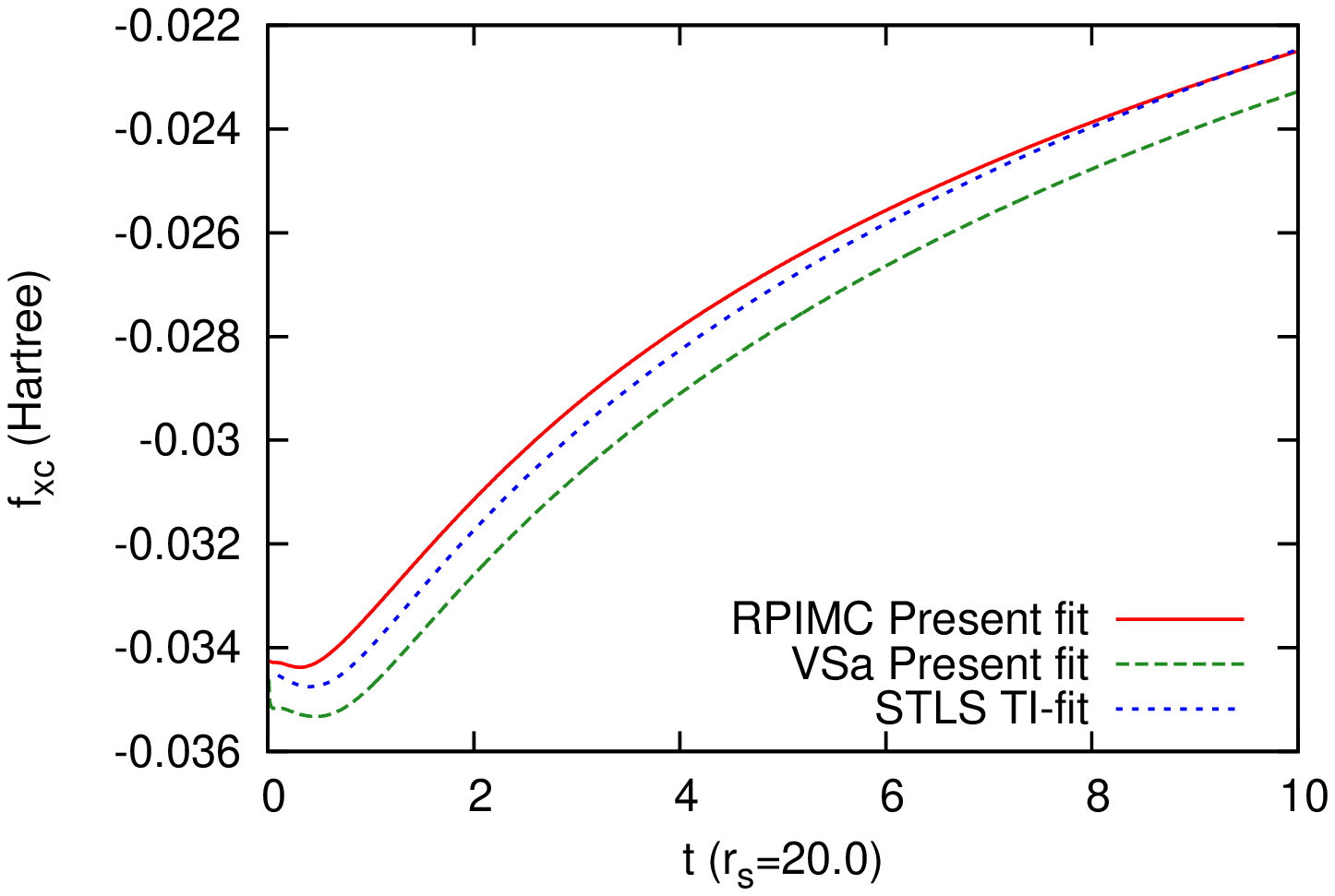}
  \caption{$f_{xc}$ calculated from the fits for STLS, RPIMC, and VSa. CHNC is also included for some.}
\end{figure*}


\begin{thebibliography}{99}
\bibitem{Lindhard} J. Lindhard, K. Dan. Vidensk. Selsk. Mat. Fys. Medd. 28,
8 (1954).

\bibitem{Gell-MannBrueckner} M. Gell-Mann and K.A. Brueckner, Phys. Rev.
106, 364 (1957).

\bibitem{STLS} K.S. Singwi, M.P. Tosi, R.H. Land, and A. Sj\"olander, Phys.
Rev. 176, 589 (1968).

\bibitem{VS} P. Vashista and K.S. Singwi, Phys. Rev. B 6, 875 (1972).

\bibitem{CA80} D.M Ceperley and B.J. Alder, Phys. Rev. Lett. 45, 566 (1980).

\bibitem{PZ82} J.P. Perdew and A. Zunger, Phys. Rev. B 23, 5048 (1981).

\bibitem{Brownetal} E.W. Brown, B.K. Clark, J.L. DuBois, and D.M. Ceperley,
Phys. Rev. Lett. 110, 146405 (2013); arXiv:1211.6130
[cond-mat.str-el] (2012). Data fit is given in 
E.W. Brown, J.L. DuBois, M. Holzmann, and D.M. Ceperley,Phys. Rev. B 88,
081102(R) (2013).

\bibitem{GuptaRajagopal} U. Gupta and A.K. Rajagopal, Phys. Rev. A 22, 2792
(1980).

\bibitem{PDw84} F. Perrot and M.W.C. Dharma-wardana, Phys. Rev. A 30, 2619
(1984).

\bibitem{TanakaIchimaru} S. Tanaka and S. Ichimaru, J. Phys. Soc. Jpn. 55,
2278 (1986).

\bibitem{DAC} R.G. Dandrea, N.W. Ashcroft, and A.E. Carlsson, Phys. Rev. B
34, 2097 (1986).

\bibitem{SchwengBohm} H.K. Schweng and H.M. B\"ohm, Phys. Rev. B 48, 2037
(1993).

\bibitem{StolzmannRosler} W. Stolzmann and M. R\"osler, Contrib. Plasma
Phys. 41, 203 (2001).

\bibitem{TanakaIchimaruMCA} S. Tanaka and S. Ichimaru, Phys. Rev. B 39, 1036
(1989).

\bibitem{Ebeling} W. Ebeling, Contrib. Plasma Phys. 30, 553 (1990).

\bibitem{PDw2000} F. Perrot and M.W.C. Dharma-wardana, Phys. Rev. B 62,
16536 (2000).

\bibitem{DuttaDufty} S. Dutta and J. Dufty, Phys. Rev. E 87, 032102 (2013);
Euro. Phys. Lett., 102 67005 (2013).

\bibitem {Hansenbook}J-P Hansen and I. MacDonald, \emph{Theory of Simple Liquids},
(Academic Press, London, 2006).

\bibitem{Hansen} J.P. Hansen, Phys. Rev. A 8, 3096 (1973).

\bibitem{Krempetal} D. Kremp, M. Schlanges, and W. Kraeft, \textit{Quantum
Statistics of Nonideal Plasmas} (Springer-Verlag 2005), p. 256.

\bibitem{Mahan} G.D. Mahan, \textit{Many Particle Physics: Second Edition}
(Plenum Publishing 1990), p. 467.

\bibitem{Ichimaru93} S. Ichimaru Rev. Mod. Phys. 62, 255 (1993)
\end{thebibliography}
\end{document}